\documentclass[smallcondensed]{svjour3}     % onecolumn (ditto)
\smartqed  % flush right qed marks, e.g. at end of proof

\usepackage{aas_macros}
\usepackage{amsmath,amssymb}
\usepackage{graphicx}
\usepackage{natbib}
\usepackage{bm}

\usepackage[colorlinks=true,citecolor=blue,linktocpage=true]{hyperref}
\bibpunct{(}{)}{;}{a}{}{,} %% natbib format like A&A and ApJ
\usepackage[pagewise]{lineno}
\usepackage{xspace}

\usepackage{xcolor}

\usepackage{multirow}

\newcommand{\Alf}{Alfv$\acute{\rm e}$n}

\newcounter{RomanNumber}
\newcommand{\MyRoman}[1]{\setcounter{RomanNumber}{#1}\Roman{RomanNumber}}

\def\myvec#1{\ensuremath{\bm{#1}}}

\newcommand{\bracs}[1]{\langle\overline{\mbox{#1}}\rangle}

\newcommand{\kms}{\ensuremath{{\rm km}~{\rm s}^{-1}}}
\newcommand{\megam}{\ensuremath{{\rm Mm}}} 

\newcommand{\betai}{\ensuremath{\beta_{\rm i}}}

\newcommand{\mui}{\ensuremath{\mu_{\rm i}}}

\newcommand{\cs}{\ensuremath{c_{\rm s}}}
\newcommand{\csi}{\ensuremath{c_{\rm si}}}
\newcommand{\cse}{\ensuremath{c_{\rm se}}}

\newcommand{\va}{\ensuremath{v_{\rm A}}}
\newcommand{\vai}{\ensuremath{v_{\rm Ai}}}
\newcommand{\vae}{\ensuremath{v_{\rm Ae}}}

\newcommand{\ct}{\ensuremath{c_{\rm T}}}
\newcommand{\cti}{\ensuremath{c_{\rm Ti}}}

\newcommand{\vf}{\ensuremath{v_{\rm f}}}
\newcommand{\vfi}{\ensuremath{v_{\rm fi}}}

\newcommand{\omgR}{\ensuremath{\omega_{\rm R}}}
\newcommand{\omgI}{\ensuremath{\omega_{\rm I}}}

\newcommand{\kcl}{\ensuremath{k_{{\rm c}, l}}}

\newcommand{\vph}{\ensuremath{v_{\rm ph}}}
\newcommand{\vgr}{\ensuremath{v_{\rm gr}}}
\newcommand{\rhoi}{\ensuremath{\rho_{\rm i}}}
\newcommand{\rhoe}{\ensuremath{\rho_{\rm e}}}

\newcommand{\vgmin}{\ensuremath{v_{\rm gr}^{\rm min}}}

\newcommand{\mathd}{\ensuremath{{\rm d}}}

\newcommand{\sech}{\ensuremath{{\rm sech}}}

\begin{document}

\title{Magnetohydrodynamic Fast Sausage Waves in the Solar Corona}
%\titlerunning{Short form of title}

\author{
B. Li           		\and
P. Antolin				\and 
M.-Z. Guo				\and		
A. A. Kuznetsov		 	\and 
D. J. Pascoe			\and
T. Van Doorsselaere 	\and 
S. Vasheghani Farahani 
}


\institute{
B. Li \at
   Shandong Provincial Key Laboratory of Optical Astronomy
      and Solar-Terrestrial Environment,
   Institute of Space Sciences, Shandong University, Weihai 264209, China   
      \email{bbl@sdu.edu.cn}           %  \\
           \and
P. Antolin			\at
   Department of Mathematics, Physics and Electrical Engineering, 
   Northumbria University, Newcastle upon Tyne NE1 8ST, UK  
	   %\email{}
	   		\and 
M.-Z. Guo	\at 
   Shandong Provincial Key Laboratory of Optical Astronomy
      and Solar-Terrestrial Environment,
   Institute of Space Sciences, Shandong University, Weihai 264209, China   
			\and 
A. A. Kuznetsov  \at
   Institute of Solar-Terrestrial Physics, Irkutsk 664033, Russia
      %\email{a\_kuzn@iszf.irk.ru}        
			\and 
D. J. Pascoe		\at 
   Centre for mathematical Plasma Astrophysics, 
   Mathematics Department, KU Leuven, Celestijnenlaan 200B bus 2400, 
   B-3001 Leuven, Belgium		
			\and
T. Van Doorsselaere 	\at 
   Centre for mathematical Plasma Astrophysics, 
   Mathematics Department, KU Leuven, Celestijnenlaan 200B bus 2400, 
   B-3001 Leuven, Belgium		
			\and 
S. Vasheghani Farahani \at 
   Department of Physics, 
   Tafresh University, Tafresh 39518 79611, Iran
}

\date{Received: date / Accepted: date}

\maketitle

\begin{abstract}
Characterized by cyclic axisymmetric perturbations to both 
    the magnetic and fluid parameters,
    magnetohydrodynamic fast sausage modes (FSMs) 
    have proven useful for
    solar coronal seismology given their strong dispersion. 
This review starts by summarizing the dispersive properties
    of the FSMs in the canonical configuration 
    where the equilibrium quantities are transversely structured
    in a step fashion.
With this preparation we then review the recent theoretical
    studies on coronal FSMs, showing that 
    the canonical dispersion features 
    have been better understood physically,
    and further exploited seismologically. 
In addition, we show that departures from the canonical equilibrium configuration
    have led to qualitatively different
    dispersion features, thereby substantially  
    broadening the range of observations
    that FSMs can be invoked to account for.
We also summarize the advances in forward modeling studies,
    emphasizing the intricacies in interpreting observed oscillatory
    signals in terms of FSMs.
All these advances notwithstanding, we offer a list of aspects that remain to be 
    better addressed, with the physical connection of coronal FSMs to the quasi-periodic pulsations in solar flares particularly noteworthy. 
%
%Insert your abstract here. Include keywords, PACS and mathematical
%subject classification numbers as needed.
\end{abstract}

\section{Introduction}
\label{sec_intro}

Despite the considerable advances in modern instrumentation,
    it remains challenging to directly measure some key parameters
    of the highly structured solar corona, 
    the magnetic field being particularly noteworthy~\citep[e.g.,][]{2009SSRv..144..413C}. 
Coronal seismology has proven valuable for indirectly
    probing coronal structures by combining the measurements
    of the hosted low-frequency waves with continuously refined theories of 
    magnetohydrodynamic (MHD) waves in an inhomogeneous medium
    (\citeauthor{1984ApJ...279..857R}~\citeyear{1984ApJ...279..857R},
    see also the reviews by e.g.,
    \citeauthor{2000SoPh..193..139R}~\citeyear{2000SoPh..193..139R},
    \citeauthor{2005LRSP....2....3N}~\citeyear{2005LRSP....2....3N},
	\citeauthor{2020ARAA..Nakariakov}~\citeyear{2020ARAA..Nakariakov}).
The terminology in modern coronal seismology comes primarily
    from the theoretical work by \citet[][hereafter ER83]{1983SoPh...88..179E}.
Among the many ``collective waves'' (``modes'' hereafter), sausage modes are the simplest
    in that the associated perturbations are axisymmetric, but prove theoretically  
    intriguing and seismologically important given their dispersion. 
Take fast sausage modes (FSMs) {  for instance, which will be} the focus of this review. 
Theoretically, two intriguing features arise, 
    one being that FSMs can be trapped only when their axial
    wavenumbers exceed some cutoff \citep{1984ApJ...279..857R},
    the other being that their axial group speed 
    possesses a nonmonotonic frequency dependence~\citep{1983Natur.305..688R}. 
The former feature leads to
    the periods of standing FSMs being comparable to the transverse \Alf\ time.
For impulsively excited sausage wavetrains, the two features lead to 
    a distinctive temporal profile comprising a periodic phase, a quasi-periodic phase, 
    and an Airy (decay) phase when the profile is sampled sufficiently far
    from the exciter.
Regardless, the characteristic periodicity remains similar
    to the transverse \Alf\ time, which evaluates to seconds to tens of seconds for
    typical coronal structures. 
Consequently, FSMs in flare-associated structures were proposed to interpret
    second-scale oscillations often seen 
    in type IV radio bursts  
    (\citeauthor{1983Natur.305..688R}~\citeyear{1983Natur.305..688R}, 
	see also \citeauthor{1970A&A.....9..159R}~\citeyear{1970A&A.....9..159R},
	\citeauthor{1975IGAFS..37....3Z}~\citeyear{1975IGAFS..37....3Z},
	\citeauthor{1978SoPh...58..165M}~\citeyear{1978SoPh...58..165M}
	). 
This interpretation remains popular 
    today~\citep[e.g.,][]{2013A&A...550A...1K,2018ApJ...855L..29K},
    and has been generalized to be a strong candidate mechanism
    for interpreting and seismologically exploiting
    short-period Quasi-Periodic Pulsations (QPPs) in flare lightcurves
    (see e.g., the reviews by
    \citeauthor{2009SSRv..149..119N}~\citeyear{2009SSRv..149..119N},
    \citeauthor{2016SoPh..291.3143V}~\citeyear{2016SoPh..291.3143V},
    \citeauthor{2018SSRv..214...45M}~\citeyear{2018SSRv..214...45M},
    \citeauthor{2020STP.....6a...3K}~\citeyear{2020STP.....6a...3K};
    also the review by Zimovets et al. in this issue).
Likewise, impulsively generated wavetrains have been
    generalized to account for short-period oscillatory signals associated
    with active region (AR) loops observed in white light
    and various coronal forbidden lines at total eclipses
    \citep[e.g.][]{1984SoPh...90..325P,1987SoPh..109..365P, 2001MNRAS.326..428W, 2002MNRAS.336..747W, 2003A&A...406..709K,2016SoPh..291..155S}.
    
This review is intended to summarize recent advances in the theoretical understanding
    and consequently seismological applications of coronal FSMs. 
Before proceeding, however, with AR loops and flare loops in mind, let us list some typical
    ranges of the values for the length-to-radius ratio ($L/R$), 
    plasma $\beta$, and the density contrast between the loop fluid and its surroundings.
For AR loops imaged in EUV and soft X-ray, the loop length ($L$)
     typically lies in the range of $[20, 1000]~\megam$, 
     while the half-width ($R$) is 
     $\lesssim 5~\megam$~\citep[][Figure~1]{2007ApJ...662L.119S}
     and may reach down to $\lesssim 300$~km~\citep{2017ApJ...840....4A}. 
Therefore one typically quotes $L/R \gtrsim 10-20$ for AR loops, whereas 
     $L/R$ was suggested to be smaller for flare loops~\citep{2004ApJ...600..458A}.
The value of $L/R$ for flare loops is
     actually passband-dependent, with values as small as $\lesssim 5$
     reported in radio measurements with NoRH~\citep[e.g.,][]{2013SoPh..284..559K}.      
As for the density contrast, values of $2-10$ are usually quoted for
     AR loops, while a value up to $100-1000$ is possible for
     flare loops~\citep{2004ApJ...600..458A}.
Furthermore, a value of $\beta \lesssim 1$ or even $\beta \ll 1$ is accepted 
     for coronal gases across ARs at heights $\lesssim 0.2~R_\odot$~\citep{2001SoPh..203...71G},
     and is usually quoted for AR loops as well.
Flare loops are somehow different from this overall picture of a magnetically dominated
     corona.  
In the aftermath of the impulsive energy release, the
     enhanced thermal pressure in some loop segment may be sufficiently
     dynamically important to expand the segment, leading to 
     a plasma $\beta$ of order unity therein. 
This thermal pressure enhancement may result directly from the energy release
     \citep{1982SvAL....8..132Z} or indirectly from the collisions between
     counter-flowing plasmas that in turn derive from chromospheric evaporation
     \citep[e.g.,][]{2016ApJ...833...36F,2019ApJ...877L..11R}.

Organizing the recent literature on a subject as specific as coronal FSMs
   is not easy. 
We choose to start with an in-depth
   review of the sausage modes in the ER83 equilibrium (Sect.~\ref{sec_ER83_theory}), 
   then come up with a list of primary dispersion features and 
   ask what simplifications or complications to the ER83 equilibrium
   are unlikely to qualitatively modify these features (Sect.~\ref{sec_ER83_ER83like}).
We will also overview a number of important issues that remain largely unanswered, 
    the excitation of FSMs as an example (Sect.~\ref{sec_ER83_misc}).
Section~\ref{sec_forward} summarizes the advances
    in forward modeling coronal FSMs,
    emphasizing its role for properly interpreting observations
    (Sect.~\ref{sec_forward}). 
We will then organize the recent theoretical progress by 
    addressing those modifications to the ER83 equilibrium
    that result in qualitative modifications to the dispersion features
    (Sect.~\ref{sec_nonER83}).
Some illustrative seismological applications enabled by these modifications    
    are offered in Section~\ref{sec_nonER83_seis}.
This review is summarized in Section~\ref{sec_conc}, where 
    some possible ways forward are also discussed. 
    
\section{Sausage Modes in the ER83 and ER83-like Equilibria}
\label{sec_ER83}

Unless otherwise specified, 
    this review adopts ideal MHD as the theoretical framework,
    for which the primary dependents are the mass density $\rho$, velocity $\myvec{v}$, magnetic field $\myvec{B}$, and thermal pressure $p$.
Let subscript $0$ denote the equilibrium parameters, from which 
    the plasma $\beta$ is defined by $2\mu_0 p_0/B_0^2$ with $\mu_0$
    being the magnetic permeability in free space.
The \Alf\ and adiabatic sound speeds are defined as 
    $\va = B_0/\sqrt{\mu_0\rho_0}$ and 
    $\cs = \sqrt{\gamma p_0/\rho_0}$, respectively. 
Here $\gamma=5/3$ is the adiabatic index. 
The tube speed and transverse fast speed
    are further defined by
    $\ct^2 = \cs^2 \va^2/(\cs^2+\va^2)$ and 
    $\vf^2 = \cs^2+\va^2$, respectively. 

\subsection{Theoretical Basics and Seismological Applications
    of Sausage Modes in the ER83 Equilibrium}
\label{sec_ER83_theory}    
 
ER83 modeled a coronal loop as a straight, static
      \footnote{The ambient is taken to be always static in this review.},
    density-enhanced, field-aligned tube with circular cross-section,
    enabling the natural choice of a cylindrical coordinate system $(r, \theta, z)$.
All equilibrium parameters 
    depend on $r$ in a step (piece-wise constant) fashion.
Standard eigen-mode analysis starts by Fourier-decomposing
    any perturbation as
\begin{equation}
\label{eq_Fourier_ansatz}
   \delta f(r, \theta, z;t) = 
   {\rm Re}\left\{
   \tilde{f}(r)\exp\left[-i\left(\omega t-kz-m\theta\right)\right]
\right\}~,
\end{equation}      
    where $\omega$, $k$, and $m$ represent the angular frequency, axial wavenumber,
    and azimuthal wavenumber, respectively. 
With tilde we denote the Fourier amplitude.      
This equilibrium allows both incompressible (\Alf) 
    and compressible waves. 
Among the \Alf\ waves, the $m=0$ (torsional) and $m=1$ ones are relevant.    
Among the compressible waves, relevant are
    the kink ($m=1$) and sausage modes ($m=0$)
    \footnote{
    Kink modes are addressed only when necessary in this review.
    For more details, see the companion reviews by 
    Nakariakov et al. and Van Doorsselaere et al. in this issue.
    }.
We take $k$ as real-valued, but allow $\omega$ to be complex-valued. 
If some quantity is complex, we denote
    its real (imaginary) part with subscript R (I).
The period $P$ (damping time $\tau$)
    follows from $P = 2\pi/\omgR$ ($\tau = 1/|\omgI|$).
Instabilities are not of interest, hence $\omgI \le 0$.    
The axial phase and group speeds are defined by 
    $\vph = \omega/k$ and $\vgr = \mathd \omega/\mathd k$, respectively.
For standing modes in a loop with length $L$, the axial wavenumber
    $k$ is quantized ($k = n\pi/L, n=1, 2, \cdots$).  
We follow the convention that $n=1$ represents the axial fundamental, with 
    $n \ge 2$ representing its $(n-1)$-th harmonic.  

Let the subscripts ${\rm i}$ and ${\rm e}$ denote
    the equilibrium values at the loop axis and far from the loop, respectively
    \footnote{  
    In the ER83 setup, the equilibrium quantities are transversely structured 
        in a piece-wise constant manner, meaning that
        the subscript ${\rm i}$ (${\rm e}$) applies to the entire interior
        (exterior).    
    We choose to introduce the subscripts this way such that we can avoid
        re-introducing them when discussing the equilibria with continuous transverse structuring.  
    }. 
As in ER83, by ``coronal conditions'' we refer to 
    the ordering $\cse < \csi <\vai <\vae$.
The dispersion relation (DR) for sausage modes reads
\begin{equation}
			\label{eq_DR_tophat}
\displaystyle
  \frac{\rhoi J_0(\mu_{\rm i}R)(\omega^2 - k^2 \vai^2)}
  {\mu_{\rm i} J_1(\mu_{\rm i}R)}
= \frac{\rho_{\rm e} H^{(1)}_0(\mu_{\rm e}R)(\omega^2-k^2 v_{\rm Ae}^2)}
  {\mu_{\rm e}H^{(1)}_1(\mu_{\rm e}R)}~,
\end{equation}
     where 
\begin{equation}
\label{eq_def_mu}
\displaystyle
  \mu_{\rm i, e}^2 
= \frac{(\omega^2 - k^2v_{\rm Ai, e}^2)(\omega^2 - k^2c_{\rm si, e}^2)}
       {(c_{\rm si, e}^2+v_{\rm Ai, e}^2)(\omega^2 - k^2c_{\rm Ti, e}^2)}~,
\end{equation}
     and $J_n$ and $H_n^{(1)}$ are the $n$-th-order Bessel
     and Hankel functions of the first kind, respectively (here $n = 0, 1$).
This DR was independently derived by \citet{1975IGAFS..37....3Z},
    \citet{1982SoPh...75....3S}, 
    and \citet{1986SoPh..103..277C}.
It differs in form from the independent ER83 result in that, 
    while ER83 required the perturbations be evanescent in the ambient, 
    Equation~\eqref{eq_DR_tophat} allows the perturbations
    to propagate outward.
Trapped modes arise in the former case, 
    {  where $\mu_{\rm e}^2 <0$ and $\omega$ is real}.
Leaky modes arise in the latter case, where $\omgI$ does not vanish. 
    
\begin{figure}
\centering
\includegraphics[width=0.6\columnwidth]{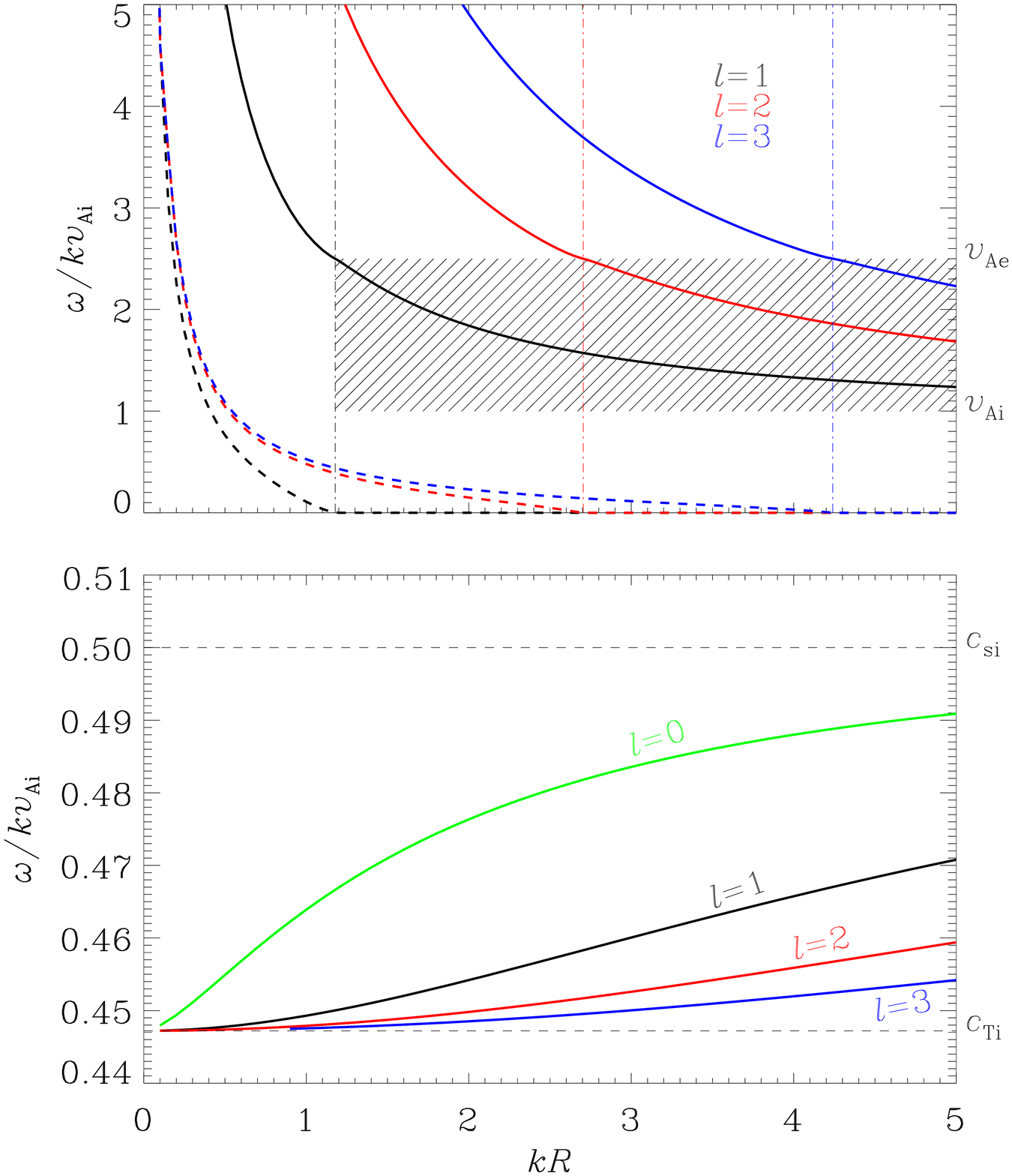}
 \caption{
 Dependence on the axial wavenumber $k$ of the axial phase speed ($\omega/k$)
     for (a) fast and (b) slow sausage modes
     in the cylindrical equilibrium considered by \citet[][ER83]{1983SoPh...88..179E}.
 Here $[\csi, \cse, \vae] = [0.5, 0.25, 2.5]~\vai$, where $\cs$ ($\va$)
     is the adiabatic sound (\Alf) speed, and the subscript i (e)
     denotes the values at the loop axis (far from the loop).
 This ordering of the characteristic speeds is typical of coronal conditions.      
 The solid and dashed curves derive from the real ($\omgR$) and imaginary ($\omgI$)
     parts of the angular frequency. 
 Note that $\omgI$ is identically zero for trapped modes, and $-\omgI$ is employed 
     to produce the dashed curves in Figure~\ref{fig_DR_tophat_vph}a.
 Slow modes are trapped regardless of $k$, whereas fast modes are trapped
     only when $k$ exceeds some cutoff as seen in the 
     hatched area.
 This figure is essentially a combination of Figure~4 in ER83, 
     Figure~1 in \citet{2007AstL...33..706K}, and Figure~6 in \citet{2016SoPh..291.3143V}.     
}
 \label{fig_DR_tophat_vph}
\end{figure}

Figure~\ref{fig_DR_tophat_vph} presents the dependence
    of the axial phase speed $\omega/k$ on the axial wavenumber $k$
    as found by solving the DR (Equation~\ref{eq_DR_tophat}).
For illustration purposes, here the characteristic speeds are specified
    as $[\csi, \cse, \vae] = [0.5, 0.25, 2.5]~\vai$.
The real and imaginary parts are plotted by the solid and dashed curves, respectively.
For completeness, slow sausage modes (SSMs) are shown
    (Figure~\ref{fig_DR_tophat_vph}b)
    in addition to FSMs (Figure~\ref{fig_DR_tophat_vph}a), the purpose being to
    show that there are an infinite number of branches, labeled by
    the transverse order $l$ and plotted by the different colors for both families. 
The smaller an $l$ is, 
    the simpler the spatial distribution of the eigen-functions.
SSMs are trapped regardless of $k$ or $l$, whereas FSMs are trapped only when 
    $k$ exceeds some $l$-dependent cutoff, 
    pertaining to the shaded area in Figure~\ref{fig_DR_tophat_vph}a. 
In addition, $\mui^2 >0$ for SSMs and trapped FSMs alike, enabling ER83 to classify them
    as ``body modes".
The peculiar label $l=0$ for SSMs, not present for FSMs, 
    is for mathematical reasons.              
When $l \ge 1$, for the $l$-th FSM,
    $\mui R$ increases monotonically from $j_{0, l}$ at the cutoff
    to $j_{1, l}$ for $kR \gg 1$, with
    $j_{n, l}$ representing the $l$-th zero 
    of the Bessel function $J_n$
    \footnote{
        	The first several zeros are
        	$j_{0, 1} = 2.4048$, $j_{0, 2} = 5.5201$,
        	$j_{1, 1} = 3.8317$, and $j_{1, 2} = 7.0156$.
        }.
For the $l$-th SSM, on the other hand, $\mu_{\rm i} R \approx j_{1, l}$ 
    in the thin-tube limit ($kR \ll 1$).      
For the peculiar $l=0$ branch of SSMs, 
    $\mu_{\rm i}$ for $kR \ll 1$ weakly depends on $kR$
    through the appearance of $\ln (kR)$, making its $k$-dependence of $\omega/k$ 
    different from the rest. 

The reason for us to make a digression to coronal SSMs, 
    the only digression in this review, is to make a better connection
    to the companion reviews on slow waves
    (Banerjee et al. and Wang et al.)
     in this topical issue.
Let $\Delta \rho$ denote the magnitude of 
    the internal density perturbation in units of
    the internal density $\rhoi$.
Furthermore, let $\Delta R$ denote the magnitude of the
    Lagrangian displacement at the loop boundary in units of
    the loop radius $R$.    
Likewise, let $\Delta v_z$ and $\Delta v_r$ denote the magnitudes of
    the axial and radial speeds, respectively.      
Extending the analysis in \citet{2016RAA....16...92Y}, it can be shown that 
    $\Delta\rho/\Delta R > 4/(\gamma \betai) = 2.4/\betai$ 
    for the branch labeled $l=0$, while taking extremely large values for $l \ge 1$.
In addition, $\Delta v_z/\Delta v_r$ exceeds
    $2[1+2/(\gamma \betai)]/(kR)$ for $l=0$, and is even larger for $l \ge 1$.
Given that $kR \ll 1$ holds in typical observations of coronal SSMs, 
    this means that it is extremely difficult
    to discern the associated expansion or contraction of the loop
    in imaging observations.
Likewise, in spectroscopic measurements, the oscillatory behavior in Doppler shift
    in general derives from the periodic variations in the axial flow.     
Physically, it is in general
    justified to see slow sausage modes as field-guided sound waves. 
This substantially simplifies the relevant efforts for modeling
    both the slow waves
    themselves~\citep[e.g.,][and references therein]{2019ApJ...886....2W}
    and 
    their observational signatures~\citep[e.g.,][]{2008SoPh..252..101D, 2009A&A...494..339O}.
On this latter aspect, we note that a proper forward modeling
    incorporating the multi-dimensional distributions of the perturbations
    is necessary when $\betai$ is not that small as happens for,
    say, flare loops~\citep{2015ApJ...807...98Y}.

For FSMs, \citet{1984ApJ...279..857R} showed that
    the cutoff wavenumber ($\kcl$) is given by 
\begin{equation}
    \label{eq_kcl_tophat}
\displaystyle
\kcl R = j_{0,l}\sqrt{
	\frac{(\csi^2+\vai^2)(\vae^2-\cti^2)}
  	     {(\vae^2-\csi^2)(\vae^2-\vai^2)}}~.
\end{equation}
For any $l$, when $k$ crosses $\kcl$ from right to left, 
    $\omgI$ is switched on, accompanied by some variation of
    the $k$-dependence of $\omgR$.
This behavior can be addressed semi-analytically but the expressions are too lengthy 
    to include~\citep{2014ApJ...781...92V}.
With $k$ further decreasing, $\omgR/k$ and $\omgI/k$ diverge 
    in a manner to ensure that $\omega$ is finite at $kR = 0$
    \citep[e.g.,][]{2007AstL...33..706K}
    \footnote{
    As detailed in Section~\ref{sec_ER83_ER83like}, much can be learned 
       for the dispersive properties of FSMs
       by examining the slab counterpart of the ER83 equilibrium.
    In the slab case, the analytical study by \citet{2019ApJ...886..112K} 
       showed that leaky FSMs of different transverse orders ($l$)
       behave differently with decreasing $k$.
    Relative to its transverse harmonics ($l \ge 2$), 
       the transverse fundamental ($l = 1$) is characterized by a 
       a faster decrease in $\omgR$ accompanied by a more rapid increase of 
       $\omgI$.
    }.
Overall, $\omgR$ decreases and hence the period $P$ increases
    with decreasing $k$ in the trapped regime.
When $k$ further decreases, the damping time $\tau$ becomes finite but 
    $P$ only weakly increases, both approaching some finite value at $kR \to 0$, 
\begin{equation}
    \label{eq_tophat_Ptau_k0}
P \approx \frac{2 \pi }{j_{0, l}}
          \frac{R}{\vfi}~,~~~
\frac{\tau}{P} \approx \frac{1}{\pi^2} \frac{\rhoi}{\rhoe}~,          
\end{equation}               
    where $\vfi$ is the internal transverse fast
    speed.
Equation~\eqref{eq_tophat_Ptau_k0} applies when $\rhoi/\rhoe \gg 1$,
    and explicitly shows that the periods of FSMs are of the 
    order of the transverse \Alf\ time.
Without further refinement, 
    Equation~\eqref{eq_tophat_Ptau_k0}
    is already seismologically useful.
Take the Culgoora radio spectrograph measurements of
    a regular series of damped pulses presented by~\citet{1973SoPh...32..485M}.
From Figure~2 therein one discerns a period $P \approx 4.3$~sec, and an e-fold damping time
    $\tau \approx 10 P$.
If attributing this rapid oscillation to a standing FSM of transverse order $l=1$
   in a long loop,
   then one deduces from Equation~\eqref{eq_tophat_Ptau_k0}
   a density contrast $\rhoi/\rhoe \approx 100$
   and a transverse fast time $R/\vfi \approx 1.6$~sec
   \citep[e.g.,][]{2007AstL...33..706K}
   \footnote{
   In the slab case, it is possible to analytically establish that 
      the damping time experiences its shortest value at the long wave-length limit when the density ratio of the internal and external medium is close to unity, while experiencing its highest value just below the cut-off value when the density ratio possesses its highest value \citep{2019ApJ...886..112K}.
   }.

\begin{figure}
\centering
\includegraphics[width=0.9\columnwidth]{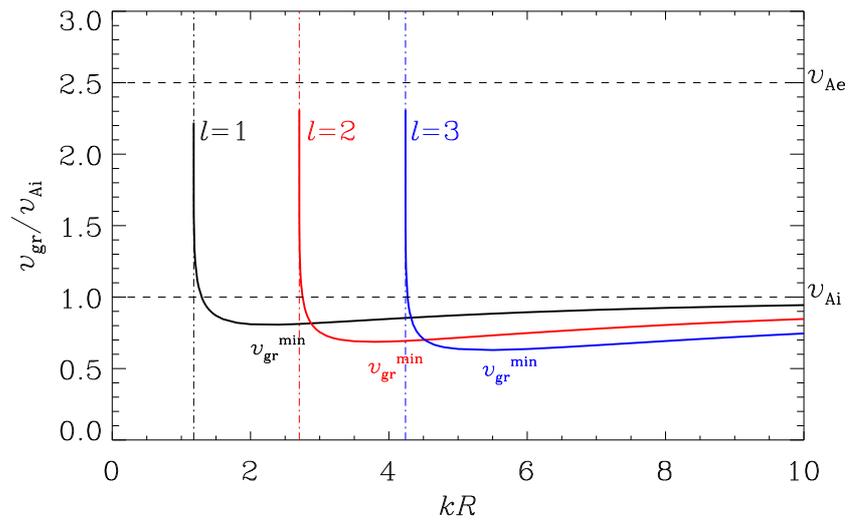}
 \caption{
  Dependence on the axial wavenumber $k$ of the axial group speed ($\vgr$)
      of trapped coronal fast sausage modes. 
  Derived from Figure~\ref{fig_DR_tophat_vph}a,
      this figure is a generalization
      of Figure~3b in \citet{1984ApJ...279..857R}.
}
 \label{fig_DR_tophat_vgr}
\end{figure}

Another dispersive feature that characterizes the ER83 equilibrium is 
    the non-monotonic frequency-dependence of the axial group speed ($\vgr$)
    of trapped FSMs, as shown in Figure~\ref{fig_DR_tophat_vgr}.
For any transverse order, $\vgr$ decreases from $\vae$ at the cutoff
    to some minimum $\vgmin$ before approaching $\vai$ from below.     
Note that this figure derives directly from Figure~\ref{fig_DR_tophat_vph},
    and that 
    the non-monotonic dependence of $\vgr$ on the axial wavenumber $k$
    is equivalent to the frequency dependence because $\omega$ increases
    monotonically with $k$. 
This feature is most relevant for discussing impulsively excited FSMs, for which
    the perturbations like, say, the Lagrangian displacement $\xi(r, z; t)$
     necessarily involve all frequencies and all wavenumbers~\citep{1984ApJ...279..857R, 1986NASCP2449..347E}. 
While initially examined by drawing analogy with the Pekeris waves in oceanography,
    this problem was recently placed on a firmer mathematical ground by
    \citet{2015ApJ...806...56O}.
First expressing $\xi(r, z; t)$ as the Fourier integral over all $k$, this
    study shows that the Fourier component $\tilde{\xi}(r, k; t)$ can be expressed as
    (see Equation~25 therein)
\begin{linenomath}
\begin{align}
   \label{eq_impl_Oliver25}       
  \tilde{\xi}(r, k; t) 
& = \sum_{l=1}^{N}
  \left[A_l^+(k) {\mathrm e}^{- i \omega_l (k) t}
       +A_l^-(k) {\mathrm e}^{i   \omega_l (k) t}\right] \hat{\xi}_l(r, k) \nonumber \\
& + \int_{|k|\vae}^{\infty}
  \left[A_\omega^+(k) {\mathrm e}^{- i \omega (k) t}
       +A_\omega^-(k) {\mathrm e}^{i   \omega (k) t}\right] \hat{\xi}_\omega(r, k){\mathd}\omega~.
\end{align}      
\end{linenomath}    
Here the superscripts $\pm$ pertain to forward
    and backward propagations in the axial direction.
For a given $k$, the summation in Equation~\eqref{eq_impl_Oliver25} incorporates the
    contribution from the $N$ discrete trapped modes with $N$ possibly zero 
    (see Figure~\ref{fig_DR_tophat_vph}).
Outside the trapped regime, the integral sees the perturbations as comprising 
    a continuum of improper modes rather than directly the discrete leaky modes.
The coefficients $A^{\pm}$ reflects how the energy contained in the
    initial perturbation is distributed among the trapped
    and continuum modes. 
When some signal $\phi$ is sampled along the axis and at distances sufficiently far from
    the exciter, trapped modes are relevant and $\phi$
    goes like~\citep[][Equation~6]{1986NASCP2449..347E}
\begin{equation}
   \label{eq_impl_ER86}
\phi(z, t) \sim \left\{ 
\begin{array}{ll}
A(z, t) {\mathrm e}^{i [k z- \omega(k) t)]}~,~~~ 
  & 	\omega''(k) \ne 0~, \\
\displaystyle 
\frac{1}{[t |\omega'''(k)|]^{1/3}}{\mathrm e}^{i [k z- \omega(k) t)]}~,~~~ 
  & 	\omega''(k) = 0~.
\end{array}
\right.
\end{equation}              
Here $' \equiv {\mathd}/{\mathd k}$, and as such Equation~\eqref{eq_impl_ER86}
   directly shows why the non-monotonic $\omega$-dependence of $\vgr = \omega'(k)$ 
   matters.
Note that $k$ is interpreted as dependent on $(z, t)$
   through $\omega'(k) = z/t$.

Equation~\eqref{eq_impl_ER86} proves theoretically insightful and seismologically useful. 
At a distance $h$ from the exciter, it offers a heuristic
   interpretation of $\phi(h, t)$ envisaged by \citet{1984ApJ...279..857R}.
When $h/v_{\rm Ae} < t < h/v_{\rm Ai}$, individual wavepackets
    with progressively low group speeds arrive consecutively, constituting
    the ``periodic phase''.
When $h/v_{\rm Ai} < t < h/\vgmin$ (see Figure~\ref{fig_DR_tophat_vgr}), 
   two wavepackets with
    the same group speed but different frequencies arrive simultaneously,
    enhancing the signal in this ``quasi-periodic phase''.
In these phases, the situation $\omega''(k) \ne 0$ arises.    
When $t> h/\vgmin$, no incoming wavepackets are expected, 
    resulting in a decay 
    phase which pertains to $\omega''(k) =0$.
With a rapid pulsation measured with the Daedalus radio spectragraph, 
    \citet{1984ApJ...279..857R} illustrated how to deduce the geometrical parameters
    ($R$ and $h$) by combining the measurements of such timescales
    as the period in 
    the decay phase and the duration of the quasi-periodic phase.

\subsection{Fast Sausage Modes in ER83-like Equilibria}
\label{sec_ER83_ER83like}    
By ``ER83-like'' we mean a rather general set of equilibria that deviate 
    from the ER83 equilibrium only slightly such that 
    the primary dispersive properties of FSMs are preserved.
The purpose is to leave more significant deviations till later, 
    and focus here on what simplifications or complications
    are unlikely to cause qualitative difference to FSMs.
By ``primary'' we list a number of dispersion features, the first two 
    not restricted to sausage modes.
\begin{enumerate}
\item The dispersive properties do not depend on the direction of propagation,
      as ensured by the absence of an equilibrium loop flow.
\item A Fourier decomposition in the axial direction is enabled by the axial homogeneity. 
\item Sausage modes can be told apart from the rest as ensured
      by sufficient symmetry.
\item All branches of FSMs possess cutoff wavenumbers regardless of the
      transverse order.
\item All branches of FSMs
      are characterized by a nonmonotical frequency dependence
      of the axial group speed in the trapped regime.                                    
\end{enumerate}    
              
As an example, one may replace the step profile in ER83 with a configuration
    comprising a uniform interior, a uniform exterior, and a transition layer (TL)
    continuously connecting the two.
In this case, \citet{2016ApJ...833..114C} showed that Feature~4 persists.
In addition, varying $\betai$ between $\sim 0$ and unity led only to
    a rather weak $\betai$-dependence for both the cutoff wavenumbers and
    the eigen-frequencies at sufficiently small $k$, provided that the transverse fast time
    ($R/\vfi$) is used to measure the timescales.
It follows that a finite plasma $\beta$ is unlikely to alter
   the dispersive properties found with $\beta=0$.
Indeed, the zero-$\beta$ study examining the $\vgr-\omega$
   curves indicated that Feature 5 persists, despite that multiple extrema may appear  for thick TLs~\citep{2016ApJ...833...51Y}.
Seismologically, if the $\beta$-insensitivity always holds, then one
   is allowed to deduce the transverse \Alf\ time ($R/\vai$)
   with the simpler zero-$\beta$ MHD~\citep[e.g.,][]{2012ApJ...761..134N,2014A&A...572A..60L,2017ApJ...836....1Y,2019MNRAS.488..660L}.
It is just that this deduced transverse \Alf\ time should be interpreted 
   as actually being the transverse fast time. 

Another school of ER83-like equilibria is realized by placing
   a straight, field-aligned, density-enhanced slab in an ambient corona, as first
   considered by \citet{1982SoPh...76..239E}.
As long as the symmetry about the slab axis is maintained, 
   the dispersive properties of FSMs
   are strikingly similar to the cylindrical case.
Take step profiles.   
The cutoff wavenumber $\kcl$ is identical in form 
   to Equation~\eqref{eq_kcl_tophat} except that $j_{0,l}$ is
   {  replaced with $(l-1/2) \pi$ \citep{1995SoPh..159..213N}.
Likewise, revising $j_{0,l}$ to $(l-1/2) \pi$ 
   in Equation~\eqref{eq_tophat_Ptau_k0} yields}
   the period $P$ for small $k$
   (\citeauthor{2018ApJ...855...47C}~\citeyear{2018ApJ...855...47C};
   see also \citeauthor{2005A&A...441..371T}~\citeyear{2005A&A...441..371T})
\footnote{These two papers found that the damping time $\tau$
   (or equivalently $\omgI$) at $kR \to 0$ does not depend on the transverse order.
   In addition, $\tau/P$ is proportional to $\sqrt{\rhoi/\rhoe}$ rather than 
       $\rhoi/\rhoe$ for sufficiently large density contrasts. 
   These differences from the cylindrical case are not regarded as ``primary", though. }.     
In addition, both $\kcl$ and the eigen-frequencies at sufficiently small $k$
   are found to depend on $\betai$ only weakly, for step~\citep{2009A&A...503..569I}
   and multi-layered profiles alike~\citep{2018ApJ...855...47C}.     
As in the cylindrical case, this $\beta$-insensitivity justifies the
   rather extensive use of the
   simpler zero-$\beta$ MHD for examining coronal FSMs
   \citep[e.g.,][]{2005A&A...441..371T,2014A&A...567A..24H,2015ApJ...801...23L, 2015ApJ...814...60Y,2019ApJ...886..112K}.
However, we note by passing that if an equilibrium
   is not symmetric about the slab axis,
   then the sausage modes will be coupled to the kink ones,
   leading to a less clear distinction between the two.
{  
This asymmetry in the equilibrium arises in a number of
   observationally relevant situations.
For instance, it appears in the presence of an external rigid boundary,
   which in turn appears when one addresses an external wave source 
   \citep{2020MNRAS.496.3035L}.
Likewise, it shows up when    
   the equilibrium quantities on one side of the slab are different
   from those on the other side
   \citep[e.g.,][]{2017SoPh..292...35A,2018ApJ...853..136K,2020ApJ...890..109O}.
}

Regarding FSMs, the similarity between the slab and cylindrical geometries enables
   a unified discussion on when Features 4 and 5 occur.
This was done in zero-$\beta$ MHD for the equilibria where
   the magnetic field is uniform but the density 
   takes a generic variation 
\begin{equation}
 {\rho}(r)= \rhoe+(\rhoi-\rhoe) f(r)~.
    \label{eq_rho_profile_general}
\end{equation}
Here $f(r)$ attains unity at the loop axis ($r=0$) and zero
   when $r\to \infty$.
In the slab geometry, $r$ is interpreted as the distance ($x$) from the slab axis. 
In both geometries, cutoff wavenumbers exist only when $f$ decreases 
   more rapidly than $r^{-2}$ 
   at large distances, a concrete result obtained    
   with Kneser's oscillation theorem by 
   \citet{2015ApJ...801...23L} and \citet{2015ApJ...810...87L}.
Note that $f(r)$ varies as $r^{-\infty}$ for a uniform ambient as happens for ER83.  
On the other hand,  in general $1-f(r)$ behaves as $\propto r^{\nu}$
   close to the loop axis.
It was shown that the $\omega$-dependence of the axial group speed ($\vgr$)
   of trapped FSMs is non-monotonic when $\nu>2$
   for the cylindrical~\citep{2017ApJ...836....1Y}
   and slab geometries alike~\citep{2018ApJ...855...53L}.
Note that $\nu = \infty$ for step profiles as considered by ER83.      
Note further that such a critical $\nu$ 
   was implied in \citet{1995SoPh..159..399N}
   and \citet{2015ApJ...810...87L}. 
In the former slab study, the authors examined a generalized symmetric Epstein profile 
   $f(x) = \sech^{2}[(|x|/d)^\alpha]$ with $d$ some mean half-width of the slab,
   finding that the $\vgr-\omega$ curve is nonmonotical
   only when the steepness parameter $\alpha >1$.
This is understandable because in this case
   $f(x) \approx 1-(x/d)^{2\alpha}$ for $0 < x/d \ll 1$, resulting in a
   $\nu = 2\alpha >2$ when $\alpha >1$. 

\begin{figure}
\centering
\includegraphics[width=0.9\columnwidth]{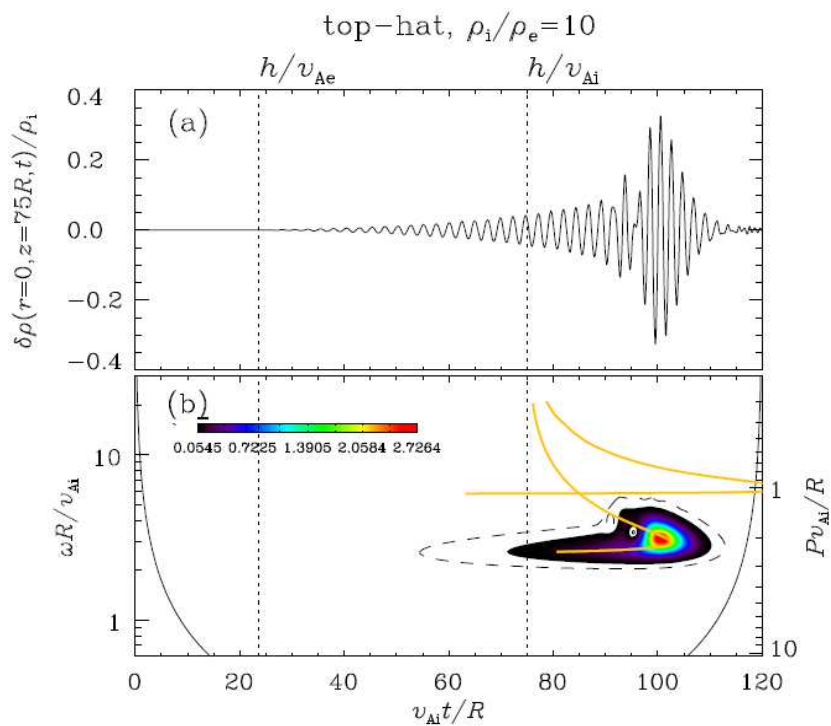}
 \caption{
 Signatures of an impulsively generated wavetrain sampled 
     at a distance $h=75R$ from the exciter along the loop axis. 
 A zero-$\beta$ version of the ER83 equilibrium is considered,
     with the density contrast being $\rhoi/\rhoe = 10$.     
 Presented are (a) the density perturbation $\delta\rho$ 
     and (b) its Morlet spectrum. 
 The Morlet spectrum from the time-dependent simulation is closely shaped by 
     the yellow curve representing $\omega-h/\vgr$ found with
     an independent eigenmode analysis.
 Taken from \citet{2017ApJ...836....1Y} with permission.    
}
 \label{fig_impl_Yu17_fig3}
\end{figure}

Now we can proceed further with impulsively generated sausage wavetrains. 
While the physics in Equations~\eqref{eq_impl_Oliver25} and \eqref{eq_impl_ER86}
   remains untouched, it seems more straightforward to work with 
   the wavelet spectra of the wavetrains rather than the time series itself
   from the seismological perspective.   
This is illustrated with Figure~\ref{fig_impl_Yu17_fig3}, 
   where the time series of the density variation
   $\delta \rho(h, t)$ and its Morlet spectrum are shown for a wavetrain 
   sampled at a distance of $h=75~R$ along the axis
   from the impulsive localized driver. 
This figure was taken from \citet{2017ApJ...836....1Y} who worked in zero-$\beta$ MHD
   and adopted a cylindrical geometry.    
Furthermore, it pertains to the simplest case where the density
   is piecewise constant with a density contrast $\rhoi/\rhoe = 10$.
As outlined by \citet{1984ApJ...279..857R}, 
   the duration of the quasi-periodic phase proves seismologically useful. 
With Equation~\eqref{eq_impl_ER86}, this quasi-periodic phase is
   expected to start at $h/\vai$ with the simultaneous arrivals of wavepackets
   with $\vgr$ lying between $\vai$ and $\vgmin$.
However, Figure~\ref{fig_impl_Yu17_fig3}a suggests that,
   without the vertical line marking $h/\vai$,
   it seems difficult to tell when the quasi-periodic phase starts
\footnote{
   This is not to say that Figure~\ref{fig_impl_Yu17_fig3} contradicts 
       Figure~3a in \citet{1984ApJ...279..857R}, which sketches the three-phase
       scenario for impulsively generated wavetrains.
   It is just that there are certain requirements for the sketch to apply, 
       with the need to sample a wavetrain at a distance sufficiently 
       far from the exciter already implied in Equation~\eqref{eq_impl_ER86}.
   On top of that, some further time-dependent simulations have suggested that 
       the duration of the exciter needs
       to be $\lesssim R/\vai$ \citep{2019A&A...624L...4G},
       and the spatial extent of the exciter needs to be comparable
       to the waveguide radius \citep{2017ApJ...836....1Y}.  
   We take these requirements as encouraging rather than discouraging
       for one to further examine
       impulsive wavetrains, because the deviations of the signatures of 
       observed wavetrains from the sketch 
       actually encode a rich set of seismic information on, say, the exciters.
   For more details, see Section~\ref{sec_seis_impls}.
}.
While this difficulty persists in Figure~\ref{fig_impl_Yu17_fig3}b,
   the Morlet spectrum 
   seems more informative in that it seems to be closely shaped by the yellow curves
   representing the wavepacket arrival time $h/\vgr(\omega)$
   as a function of $\omega$. 
Physically, this behavior is expected with Equation~\eqref{eq_impl_ER86}
   given the exponential term therein.
Seismologically, it should be useful if one can further localize the Morlet spectrum
   in the frequency direction, a point we will return to later. 
       
First found in a slab geometry by \citet{2004MNRAS.349..705N},
   the Morlet spectra similar to the one in Figure~\ref{fig_impl_Yu17_fig3}b
   are named ``crazy tadpoles" given their shape.
Since then crazy tadpoles have been found in multiple linear MHD computations 
   for both cylindrical~\citep[e.g.,][]{2015ApJ...814..135S}
    and Cartesian configurations \citep[e.g.,][]{2013A&A...560A..97P,2014A&A...568A..20P,2018ApJ...855...53L}
    \footnote{Crazy tadpoles were found in nonlinear sausage wavetrains in density-enhanced
    slabs as well~\citep{2017ApJ...847L..21P}. 
    Nonlinear waves, however, are beyond the scope of this review.}.
This insensitivity to geometry is understandable because crazy tadpoles are expected
    as long as Features 4 and 5 persist~\citep{2017ApJ...836....1Y}.
More importantly, crazy tadpoles were observed for short-period
    wavetrains propagating in AR loops as measured, say, during the Aug 1999 total eclipse
    \citep[][Figure~4]{2003A&A...406..709K}.
At this point we note that the theoretical results in wave-guiding slabs also apply
    when current sheets are embedded, provided that the electric resistivity
    vanishes~\citep[e.g.,][]{1986GeoRL..13..373E, 1997A&A...327..377S}.
Indeed, dispersive sausage wavetrains guided by flare current sheets have been
    employed to interpret
    some fine structures in broadband type IV radio bursts~\citep[e.g.,][]{
	2013A&A...550A...1K,2016ApJ...826...78Y}
	\footnote{If the width of the current sheet (CS) is taken to vanish,
	 then the slab results directly apply~\citep[e.g.][appendix]{2011SoPh..272..119F}. 
	If the CS width is taken to be finite, 
	   then the \Alf\ speed vanishes somewhere in the CS,
	   and the characteristic timescales of FSMs tend to be comparable to 
	   the transverse sound time evaluated with the internal sound speed
	   \citep[][Figure~2]{1986GeoRL..13..373E}.  
	Impulsive wavetrains, however, remain qualitatively the 
	   same in the sense of MHD~\citep[e.g.,][]{2012A&A...537A..46J,2014ApJ...788...44M},
	   which is understandable because Features~4 and 5 persist therein.  
	We note that the Morlet spectra of the modulated radio fluxes
	   further depend on the radiation mechanism that bridges
	   the MHD results and measurables.
	The spectra may be similar across a substantial range of radio frequencies,
	   as in the first identification of tadpoles in decimetric type IV bursts
	   \citep{2009ApJ...697L.108M}.
	Their morphology may be frequency-dependent at lower radio frequencies, 
	   resulting in the ``drifting tadpoles" named by~\citet{2009A&A...502L..13M}.  
	}.    

To the list of primary dispersion features of FSMs, we need to append an item 
   that pertains to continuous transverse profiles.
This item, Feature~6 hereafter, is that an equilibrium
   is considered to be ER83-like if FSMs
   are not resonantly absorbed in the \Alf\ continuum, which arises
   when the \Alf\ speed profile ($\va$) varies continuously.
Feature~6 applies to all the above-referenced equilibria despite that 
   the axial phase speed of FSMs can indeed match $\va$ at some ``resonant location''.
Mathematically, in cylindrical geometry \citet{2013ApJ...777..158S} showed that 
   this ``resonant location" is not a genuine singularity that is required 
   for an FSM to resonantly couple to the $m=0$ \Alf\ wave.
Physically, the absence of the \Alf\ resonance
   is because the velocity ($\delta \myvec{v}$) and 
   magnetic field perturbations ($\delta \myvec{B}$)
   associated with the \Alf\ waves decouple from those in FSMs~\citep[see e.g., the review by][]{2011SSRv..158..289G}.

%        
%Furthermore, in the trapped regime, extending \citet{2016RAA....16...92Y}
%    yields that $\Delta\rho/\Delta R > j_{0, l}/J_1(j_{0, l})$, which evaluates
%    to $4.63$ for $l=1$ but is substantially larger for large $l$.
%This means that in principle, the periodic variation of the loop cross-sections
%    has some chance to be discerned in imaging observations.   
    
\subsection{Miscellaneous}
\label{sec_ER83_misc}    
This section is devoted to the aspects that are important
   for coronal FSMs but not directly connected to their   
   dispersive properties and seismological applications.
   
\subsubsection{Excitation of Coronal Fast Sausage Waves}
%Dispersive properties aside, one may question how coronal FSMs are excited
%   in the first place?
%The answers to this may be grouped into two categories, with one involving 
%   and the other not involving flaring processes. 

Coronal FSMs remain primarily invoked for interpreting short-period QPPs
   in flares~\citep[e.g., the review by][]{2018SSRv..214...45M}.
However, there are only a few studies that explicitly address how FSMs may
   result from the flaring processes.
The earliest seems to be the one by \citet{1971Natur.234..140M} who
   specifically examined what caused the regular pulses in type IV radiations 
   measured with the Culgoora radiospectrograph on Sept 27, 1969.     
A coronal shock, manifested as a type II burst preceding the pulses, proved crucial.
This shock was suggested to hit a pre-existing loop at low heights, exciting \Alf\ waves
   and consequently accelerating electrons that emit radio emissions.
The shock then hit the loop top, exciting FSMs that then
   modulate the radio emissions. 
As such the \Alf\ speed in the loop was required to exceed
   the external one substantially.
A second scenario, proposed also for pulsations in type IV radio emissions,
   was offered by \citet{1978SoPh...58..165M}.
The idea is that FSMs, an eigen-oscillation of a density-enhanced loop,
   can be resonantly amplified by fast particles (primarily protons) trapped
   and bouncing in the loop. 
With the period given by Equation~\eqref{eq_tophat_Ptau_k0}, the authors deduced
   a requirement for the speeds of the bouncing protons to reach a fraction of 
   the light speed. 
A third scenario was proposed by \citet{1982SvAL....8..132Z}, who interpreted
   second-scale pulsations of flare hard X-rays by invoking FSMs in a local magnetic trap.
Initially a flare loop segment close to the flaring site,
   this trap results when the enhanced gas pressure expands the segment.
Excited by the impulsive energy release
   \footnote{
   Implied here is that the magnetic trap forms first, and 
      FSMs are excited in this preset host 
      by some continuation of the energy release.
   The energy release is nonetheless ``impulsive'' or thrust-like, 
      by which \citet{1982SvAL....8..132Z} meant that its duration 
      is shorter than the periods of the FSMs.  
   Intuitively speaking, the expanded segment may shrink again, provided that
      the magnetic field therein is not that weak.
   FSMs may therefore develop around some new quasi-equilibrium,
      in conjunction with the response of the segment to the
       ``thrust'' imparted by the energy release. 
   Thrust-excited FSMs were numerically
         explored by \citet{2009A&A...494.1119P} for 
         preset expanded loops that are in force balance with the environment.
   However, to our knowledge, one has yet to quantitatively examine
         the excitation of FSMs during the restoration processes
         of expanded loops that are not in force balance with their ambient. 
   {  At this point, we note also that the subsecond-period oscillations recently
        detected in decimetric radio bursts by \citet{2019ApJ...872...71Y}
        actually agree with the interpretation of thrust-excited FSMs in post-reconnection loops (see Figure~11 therein).
        However, different from the \citet{1982SvAL....8..132Z} scenario is that the modulation to the radio flux was attributed to the trapping and/or acceleration
        of energetic electrons within the time-varying wavepackets rather than the
        modulation of the mirror ratio of the local trap by the time-varying FSMs}. 
   }, 
   FSMs are expected to modulate
   the flux of energetic electrons escaping from the trap by varying the mirror ratio.
A modulation in hard X-ray fluxes then follows given that these electrons eventually 
   hit the dense atmospheres at the loop footpoints. 

Another group of scenarios for exciting FSMs largely concern AR loops. 
In one scenario, FSMs were observed to be generated in situ, 
   along with kink modes, 
   due to flow collision in a loop with coronal rain \citep{2018ApJ...861L..15A}.
Numerical simulations provided support to this observed in-situ excitation
   \citep{2019A&A...626A..53P}, 
   showing that while the resulting kink modes strongly depend on the symmetry of the collision (in terms of morphology, trajectory and energy),
   the FSMs are to a large extent independent. 
Such dependency leads to mostly standing FSMs and propagating kink modes. 
Physically, one expects this mechanism to be common
   in thermally unstable loops, 
   given that the local loss of pressure therein leads to collision of counter-streaming flows \citep{2020PPCF...62a4016A}. 
Its occurrence is therefore tied with the amount of thermally unstable plasmas
   in the solar atmosphere. 
Observationally, however, more needs to be done to ascertain 
   how common this excitation mechanism is,
   even though coronal rain is ubiquitously observed
    \citep{2012ApJ...745..152A}.
We note that the flow collision mechanism can be easily
   extrapolated to the flaring scenario, where
   the strong electron beam heating at both footpoints is expected to drive strong chromospheric evaporation, thus leading to highly energetic flow collision in the corona. 
This scenario is likely to be accompanied by Kelvin-Helmholtz instabilities
   due to shear flows, particularly in the presence of asymmetric footpoint heating \citep{2016ApJ...833...36F}.

FSMs in AR loops may also be connected
   to the footpoint motions in the lower atmosphere.
This connection may be direct, as proposed by \citet{1996ApJ...472..398B}
   who linked coronal FSMs to axisymmetric, radial, footpoint motions
   due to, say, exploding granules.  
Alternatively, coronal FSMs may be indirectly linked to axisymmetric, azimuthal, footpoint
   motions, which directly drive $m=0$ \Alf\ waves. 
While propagating upward, these finite-amplitude
   mother waves may nonlinearly generate 
   coronal FSMs via the ponderomotive force and nonlinear phase-mixing
   \citep{2017ApJ...840...64S}.
As demonstrated by a recent multi-instrument observational study,   
   azimuthal motions (swirls) {  are indeed abundant} in the photosphere,
   and \Alf\ pulses with sufficient amplitude can indeed be generated and reach
   at least chromospheric levels~\citep[][and references therein]{2019NatCo..10.3504L}.
   
{  Before proceeding, let us remark that, while physically sound,
     the afore-listed excitation scenarios 
     remain largely to be tested against observations.
The reason is twofold.
On the one hand, candidate coronal FSMs have only been sporadically
    reported in the literature, making it difficult to prove or disprove 
    a candidate excitation scenario in the statistical sense.
On the other hand, a substantial fraction of the excitation scenarios have
    yet to be made more quantitative to come up with a list of tell-tale
    observational signatures.
This in turn involves two steps.
Firstly, one needs to invoke one specific scenario to 
    predict, say, the characteristic periodicities
    and variations in such primitive quantities as the magnetic field.
Secondly, one needs to translate the predicted variations in
    the primitive physical variables into measurables. 
Both steps are subject to substantial intricacies, with those in the second step
    detailed in Section~\ref{sec_forward}. 
}

\subsubsection{Damping}
Being highly compressible, FSMs are likely to be damped by non-ideal effects like
   radiative loss, electron thermal conduction, and proton viscosity.
Usually the damping due to radiative loss can be neglected, at least for ER83-like equilibria
\footnote{
   Recall that the period $P$ is of the order of the transverse \Alf\ time
      ($R/\vai$, Equation~\ref{eq_tophat_Ptau_k0}).
   Adopting the radiative loss function given by 
      \citet{1978ApJ...220..643R}, one finds that 
      the associated damping timescale $\tau_{\rm rad}$ is given by $3600 T_6/n_9$~sec when $0.56< T_6 <2$ 
      and $2200 T_6^{5/3}/n_9$~sec when $2 < T_6 <10$, where $T_6$ ($n_9$)
      is the loop temperature (electron density) in $10^6$~K ($10^9$~cm$^{-3}$). 
   Typically $\tau_{\rm rad} \gg R/\vai$.
}.
However, in general the importance of the rest
   of the {  non-ideal} mechanisms depends sensitively
   on the equilibrium parameters~\citep[e.g.,][]{2007AstL...33..706K}.
The consequence is twofold.
One, an assessment of the damping rates can usually be done
    only a posteriori,   
    given that many equilibrium parameters are not known before the seismological practice.
Two, the assessment in general needs to be conducted on a case-by-case basis, given the 
    diversity of the equilibrium parameters.
Examining the \citet{1973SoPh...32..485M} event, \citet{2007AstL...33..706K}
    concluded that non-ideal effects are orders-of-magnitude weaker than the damping due to
    lateral leakage. 
    
Lateral leakage, manifested by the non-zero $\omgI$,
    is an ideal process whereby oscillating loops lose their energy
    by continuously emitting fast waves into the ambient. 
With this picture established decades ago   
    \citep{1975IGAFS..37....3Z,1982SoPh...75....3S},
    it suffices to highlight the importance of examining leaky FSMs
    from an initial-value-problem (IVP) rather than an eigen-value-problem (EVP)
    perspective. 
The problem is, 
    the eigen solutions to the EVP problem increase indefinitely with distance
    in the exterior, making its physical reality apparently questionable.
However, this reality can be recognized in two ways, both pertaining to 
    the temporal evolution of the loop system~\citep{1986SoPh..103..277C}.
If a continuous energy supply is available, 
    then a steady state can be maintained where the system oscillates at the eigen-period but experiences no damping. 
If the waves are excited by an initial ``kick'', then outgoing fast waves result
    in the exterior, the amplitude increasing with distance behind the wave front at a given instant but diminishing with time at a given location.  
Both expectations have been extensively seen in time-dependent computations for loops
    experiencing  kicks~\citep[e.g.,][]{2005A&A...441..371T,2007SoPh..246..231T,2012ApJ...761..134N}.   
In particular, the periods and damping times agree with the
    eigen-frequencies, provided that 
    an FSM with a particular transverse order $l$ is primarily excited.
This is true for both $l=1$~\citep[e.g.,][]{2016SoPh..291..877G} and higher $l$
    \citep{2020ApJ...893...62L}. 
The EVP approach can therefore be seen as a shortcut given that
    it is usually much less computationally expensive than the IVP approach.
Physically, with the system evolution in general expressible by
    Equation~\eqref{eq_impl_Oliver25}, \citet{2015ApJ...806...56O} proposed that
    the outgoing waves result from the interference of the continuum modes.
The agreement in the periods and damping times between the EVP and IVP approaches
    is therefore likely to result from this interference as well,
    with the assertion regarding
    the period explicitly demonstrated by \citet{2007PhPl...14e2101A}.

\subsubsection{Multi-stranded vs. Monolithic Loops}
\begin{figure}
\centering
\includegraphics[width=0.6\columnwidth]{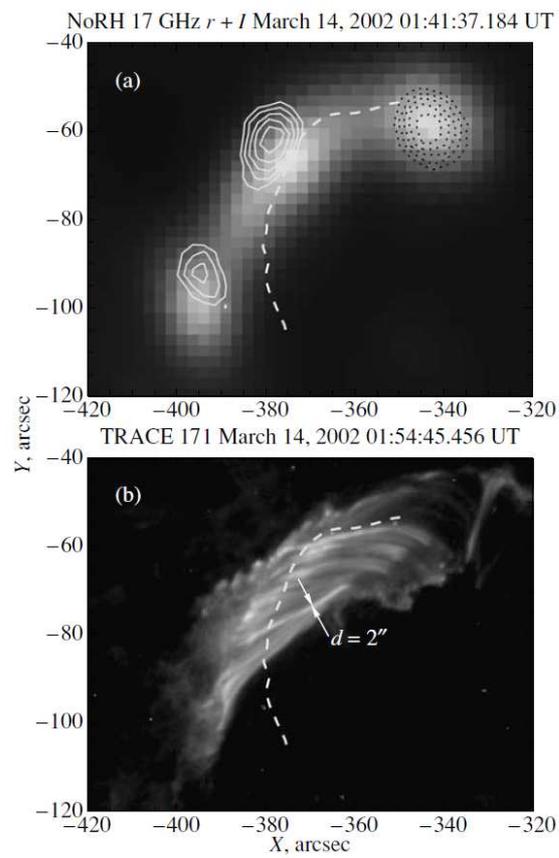}
 \caption{
 Structures associated with quasi-periodic pulsations (QPPs) as imaged in (a) radio
     and (b) EUV pertaining to an M5.7 flare on
     March 14 2002.  
 Taken from \citet{2013AstL...39..267Z}  with permission.
}
 \label{fig_zimo_multistranded}
\end{figure}

In principle, an unambiguous definition of sausage modes is ensured only 
    by the sufficient symmetry of the equilibria.
With Figure~\ref{fig_zimo_multistranded}, 
    taken from \citet{2013AstL...39..267Z}, we show that 
    in some spatially resolved measurements of QPPs in flares,
    what appears as an isolated loop in moderate-resolution images
    (acquired with, say, NoRH, Figure~\ref{fig_zimo_multistranded}a)
     may be further resolved into
    a multitude of fine strands in high-resolution images
    (observed with, e.g., TRACE, Figure~\ref{fig_zimo_multistranded}b).
Such a system does not seem to possess sufficient symmetry, and  
    a theoretical study explicitly addressing
    the existence of sausage modes in similar systems has yet to appear.
Nonetheless, some IVP studies on FSMs in
    coronal loops with fine structures in the form
    of concentric shells indicated that 
    the periods~\citep{2007SoPh..246..165P} 
    and damping times~\citep{2015SoPh..290.2231C}
    are consistent with the values found with monolithic loops, provided 
    that the number of shells exceeds, say, $\sim 10$. 
This insensitivity to fine-structuring is not an answer to the question at hand 
    because of the symmetry of the adopted equilibria.
However, the hint is that sausage-like modes are expected if the spatial scales of
    the initial perturbations considerably exceed those associated with fine structures.
By ``sausage-like" we mean the modes characterized by coherent breathing motions of
    a collection of strands as happens for an ER83-like loop
    experiencing sausage modes.
In fact, this intuitive expectation has been practiced in a considerable number
    of observational
    studies on sausage modes in the lower solar atmosphere
    \citep[e.g.,][]{2008IAUS..247..351D,2012NatCo...3.1315M,2015ApJ...806..132G},
    despite the irregularity of the shape of the examined structure 
    and the fine-structuring therein
    \citep[e.g.,][Figure~4a]{2018ApJ...857...28K}. 
    
\subsubsection{Loop Curvature}
As envisaged by \citet{2000SoPh..193..139R}, the distinction between sausage 
    and kink modes will be less clear in curved than in straight tubes, one reason
    being that kink modes will lead to the breathing motions that characterize
    sausage modes. 
A substantial number of theoretical studies on collective waves
    in curved loops then followed, the focus nonetheless being kink modes
    (see the review by~\citeauthor{2009SSRv..149..299V}~\citeyear{2009SSRv..149..299V},
    and further studies, e.g., \citeauthor{2009A&A...506..885R}~\citeyear{2009A&A...506..885R},
    \citeauthor{2011ApJ...728...87S}~\citeyear{2011ApJ...728...87S},
    and
    \citeauthor{2020ApJ...894L..23M}~\citeyear{2020ApJ...894L..23M}).
The studies explicitly addressing FSMs pertain exclusively to curved slabs
    in a zero-$\beta$ equilibrium either directly described by ~\citep[e.g.,][]{2006A&A...446.1139V,2006A&A...449..769V,2006A&A...455..709D}
    or similar to~\citep[e.g.,][]{2006A&A...456..737D,2015ApJ...814..105H,2017A&A...608A.108T}
    the following configuration. 
A cylindrical geometry $(\bar{r}, \bar{\phi}, \bar{z})$ is involved, 
    the corresponding Cartesian system being $(\bar{x}, \bar{y},\bar{z})$.
The equilibrium magnetic field $\myvec{B}_0$ is purely in the $\bar{\phi}$-direction,
    its strength $\propto 1/\bar{r}$. 
With the plane $\bar{y}=0$ representing the dense photosphere, 
    all field lines form line-tied semi-circles.
The mass density depends only on $\bar{r}$ as $\propto \bar{r}^\alpha$ with $\alpha$ being
    a constant. 
The constant of proportion in the region
    $|\bar{r}-\bar{r}_0| \le R$ is different from the rest, thereby outlining
    a density-enhanced curved slab with length $L = \pi \bar{r}_0$ and semi-width $R$.
For waves propagating only in the $\bar{x}-\bar{y}$ plane, the velocity perturbation
    is in the $\bar{r}$-direction only.     
The $\bar{\phi}$-independence of the equilibrium parameters and the line-tied boundary
    condition further dictate that the Fourier amplitude of 
    this perturbation is $\propto \tilde{v}(\bar{r}) \sin(\bar{m}\bar{\phi})$, with
    $\bar{m}=1,2,\cdots$ being the axial harmonic number.
As outlined by \citet{2006A&A...446.1139V} and \citet{2006A&A...455..709D}, 
    the distinction between sausage and kink modes
    relies on the inspection of the spatial profile of $\tilde{v}(\bar{r})$,
    the modes with breathing motions being classified as sausage modes. 
Furthermore, whether modes are leaky depends on the sign of $\mu_{\rm e}^2$ defined
    in Equation~\eqref{eq_def_mu}, with $k$ now $\bar{r}$-dependent because of
    the $\bar{r}$-dependence of the length of field lines. 
\citet{2006A&A...446.1139V} showed that 
    the modes 
    {  leak towards large (small) $\bar{r}$ when $\alpha > -4$ ($\alpha < -4$).
In the particular case where $\alpha=-4$, trapped modes arise
    when the phase speed is below the \Alf\ speed profile
    (see Figure~2 therein).    
     }
        
Of interest is how well the theoretical results on FSMs in straight slabs
    apply. 
When $\alpha=-4$, a remarkable one-to-one correspondence between the curved
    and straight cases is seen,
    and the difference in the eigen-frequencies is discernible 
    only when $L/R$ is unrealistically small~\citep[][Figures 4 and 11]{2006A&A...446.1139V}.
On the contrary, the eigen-frequencies agree with the straight-slab
    results only for sufficiently small $L/R$ when $\alpha=0$, corresponding to
    a piece-wise constant transverse profile for the density. 
For this agreement to happen, however, $L/R$ does not need to be unrealistically
    small~\citep[][Figure~7]{2006A&A...455..709D}.
In addition, while sausage modes are indeed always
    leaky in this case, the damping rate can be
    sufficiently weak to ensure observability for sufficiently large
    density contrasts (Figure~8 therein).
The same happens also for the FSMs examined in the study by \citet{2016A&A...593A..52P},
    where the equilibrium essentially pertains to
    the situation where $\alpha=-2$, 
    corresponding to a piece-wise constant transverse profile
    for the \Alf\ speed.
In this case, the periods of FSMs differ from the straight-slab expectations
    by no more than $\sim 7.5\%$ for an $L/R$ of $8.3$ and 
    a density contrast of $50$, both values being
    reasonable for flare loops
    \footnote{
    In this context, we note that impulsively generated wavetrains have also
       been modeled for curved slabs by \cite{2014A&A...569A..12N} with applications
       to AR loops in mind.
    If we extend the idea to non-straight equilibria rather
       than specifically curved loops, 
       then magnetic funnels were modeled by \cite{2013A&A...560A..97P} and coronal holes by \cite{2014A&A...568A..20P}. 
    These studies demonstrate that leaky components can form quasi-periodic wavetrains
       in the external medium.
    Unlike the simple straight slab geometry, the propagation of these wavetrains
       is no longer necessarily perpendicular to the waveguide. 
    For example, the refraction due to the nonuniform external medium 
       can allow the wavetrain to propagate in the direction of the structure,
       despite that the wavetrain itself is outside the structure
       and not structure-guided.
    }.  

\subsubsection{Energy Propagation in Sausage Modes}
Given the prominence of the coronal heating problem
   in solar and astrophysics, 
   one naturally asks how much energy is propagated in sausage modes (see also the review by Van Doorsselaere et al., this issue). 
This is particularly relevant for the lower solar atmosphere, 
   where sausage modes have been observed to be sufficiently energetic
   in a considerable number of structures of various sizes, with sunspots \citep{2014A&A...563A..12D}, 
   pores \citep{2015ApJ...806..132G}, 
   {  mottles \citep{2012NatCo...3.1315M},
   and chromospheric brightenings associated with plages/enhanced-network regions
   \citep{2020arXiv200804179G} 
   } 
   being examples.    
Theoretical examinations were therefore performed
   for sausage modes from the energetics perspective with photospheric and/or chromospheric applications in mind.
However, they are    
   expected to readily find applications to
   future high-resolution observations of sausage modes in the higher corona.
Indeed, for FSMs that are in-situ generated through flow collision,
   the observed amplitudes can reach up to $\sim 40~\kms$
   for cool an dense coronal rain flows, leading to energy flux
   densities on the order of $10^7-10^8$~erg~cm$^{-2}$~s$^{-1}$ \citep{2018ApJ...861L..15A}. 
The energy flux densities remain substantial
   for more ``typical'' coronal conditions,
   despite that they are roughly $2-3$ orders of magnitude lower 
   in this case.

To calculate the energy in sausage modes, \citet{2015A&A...578A..60M} 
   used the theory that was worked out for kink modes by \citet{2013ApJ...768..191G}. 
They computed the kinetic and magnetic energy in the ER83 model 
   with zero plasma-$\beta$. 
\citet{2015A&A...578A..60M} additionally assumed $m=0$ symmetry and
   also extended the theory to non-zero plasma-$\beta$, 
   and incorporated internal energy as well. 
However, the latter was not important for the energy flux in FSMs, 
   because their energy flux is entirely in the Poynting flux $\bracs{\myvec{S}}$. 
For FSMs with frequency near the cut-off frequency, 
   the energy flux is purely in the magnetic field direction and is given by
\begin{equation}
  \bracs{\myvec{S}}
= 2\frac{\rhoi k^2 \vai^2+ \rhoe k^2 \vae^2\ln(1/f)}{4}
\vae \pi R^2 \Xi_r^2 \hat{z},
	\end{equation}
	where $\Xi_r$ is the maximal radial displacement at the loop boundary. 
In this equation $f$ plays the role of filling factor: 
   it measures the fractional surface of density enhanced flux tubes in a cross-section of a bundle of loops 
   \citep[for more information and explanation, see][]{2014ApJ...795...18V}. 
By considering the limits $f\to 0$ (a single loop), 
   this equation shows that the exterior part is 
   causing troubles because of the leaky nature of FSMs. 
Indeed, in that limit, the Hankel functions
   in the exterior will not be square-integrable and lead to infinite energy.
For completeness, the expressions
   for energy propagation
   in the SSMs in the long-wavelength limit is given by
\begin{equation}
  \bracs{\myvec{T}} 
=  
\frac{\rhoi}{2}k^2 \cti^2 \pi R^2 \Xi_z^2 \cti \hat{z},
\end{equation}
	where $\Xi_z$ is now the longitudinal component of the displacement at the loop boundary. 
The energy flux is only in the internal energy flux, 
    is only inside the loop (and thus only determined by interior loop parameters), 
    and is once again purely along the magnetic field. 
For more general expressions and more detailed explanations
    for their derivations, see \citet{2015A&A...578A..60M}. 
These formulae were successfully used to observationally determine
    the energy content of sausage waves in e.g. \citet{2015ApJ...806..132G,2018ApJ...857...28K,2020arXiv200711594G}.

\section{Forward Modeling of Coronal Fast Sausage Modes}
\label{sec_forward}    
The fluid and magnetic parameters from analytical and numerical
    studies on coronal FSMs usually do not translate into observables
    in a straightforward manner, thereby necessitating proper forward modeling
    (see also the review by Anfinogentov et al. in this issue
    for forward modeling studies on other modes).
This can be illustrated with the intensity ($I$) of emission lines in, say, (E)UV
    for which the corona is optically thin~\citep[e.g.,][]{2018LRSP...15....5D},
\begin{equation}
   I = \int_{\rm LoS} \frac{\epsilon}{4\pi} ds~.
\end{equation}
Here $s$ is the coordinate along {  the line of sight} (LoS),
   and by integration $I$ collects all photons emitted from a square of size
    $\Delta_{\rm pixel}$ when projected onto the plane of sky (PoS). 
Furthermore, the emissivity $\epsilon$ is given by $N^2 G_{\lambda 0}$ with $N$ 
    being the electron density and $G_{\lambda 0}$ the contribution function
    for an emission line centered at $\lambda_0$ in rest frame. 
Substantial atomic physics is involved in $G_{\lambda 0}$, where all terms depend
    essentially only on the electron temperature $T$ with the possible exception 
    of the ionic fraction ($f_q$) of the line-emitting ions
    \citep[e.g.,][Equation~23]{2019ApJ...870...99S}.
In equilibrium ionization (EI), this $f_q$ is essentially only $T$-dependent as well, making
    $G_{\lambda 0}$ dependent only on $T$.
Taking advantage of the simplicity of EI, 
    \citet[][also the references therein]{2012A&A...543A..12G} were the first to recognize the instrument-sensitivity of the detectability of FSMs by further assuming 
    the $T$-profile to be uniform along an LoS. 
The result is that, if the pixel size $\Delta_{\rm pixel}$ substantially exceeds
    the loop width, then the leading-order density perturbations do not survive the
    LoS integration~(Figure~4 therein), thereby making the detection
    of FSMs unlikely.
Physically, this was attributed to that FSMs displace fluid parcels primarily
    in the transverse direction in a low-$\beta$ equilibrium
    but mass conservation is always maintained. 
        
\subsection{Thermal Emissions}
\label{sec_forward_thermal}    
Forward models that examine the modulation of thermal emissions by FSMs
   have been conducted only for emission lines in the (E)UV
   and only for ER83-like equilibria.
A notable step forward was made by \citet[][AVD13]{2013A&A...555A..74A}, who  
   incorporated $G_{\lambda 0}$ in the examinations on how trapped FSMs
   modulate the Fe \MyRoman{9}~171~\AA\ and Fe \MyRoman{12}~193~\AA\
   emissions detected with both imaging and spectroscopic instruments. 
The spatial profiles of the fluid parameters were essentially
   a three-dimensional representation of the eigen-solutions
   in an ER83 equilibrium. 
AVD13 noted that, unlike for photospheric conditions, 
   the variation in the loop radius produced by the FSM
   tends to be $\lesssim 25$~km for typical coronal conditions 
   (this results from the strong compressibility of FSMs, 
     see e.g., 
     Equation~10 in \citeauthor{2016RAA....16...92Y}~\citeyear{2016RAA....16...92Y}).
The detectability of FSMs was shown by AVD13 to
   depend sensitively on such geometrical parameters as viewing angles
   of the LoS,
   and, in particular, on the temporal and spatial resolutions
   of the instrument.
The issue with temporal cadence arises from the short period of FSMs,
   which led AVD13 to propose that the 
   ionic fractions ($f_q$) may not respond instantaneously
   to the temperature variations to maintain ionization equilibrium.
Indeed, it was shown by \citet{2019ApJ...874...87S} that
    non-equilibrium ionization (non-EI) effects are negligible only when
    the angular frequencies $\omega$ of FSMs are far below
    the ionization and recombination frequencies ($\nu_{\rm C}$ and $\nu_{\rm R}$).
This requirement is more stringent than that the wave period $P=2\pi/\omega$ be 
    much longer than the ionization and recombination timescales 
    ($1/\nu_{\rm C}$ and $1/\nu_{\rm R}$).
Non-EI effects are therefore seen even in the study on the Fe~\MyRoman{21}~1354~\AA\
    line emitted from flare loops, despite the high densities therein
    \citep{2019ApJ...874...87S}. 
That said, the non-EI computations differ from their EI counterparts only
    in intensity variations, whereas the Doppler shifts and Doppler widths
    hardly experience any influence. 
Indeed, AVD13 showed that the wave modulation of spectroscopic quantities is
    substantial, particularly for the Doppler width.    

Summing up AVD13, 
    \citet{2019ApJ...870...99S} and \citet{2019ApJ...874...87S},
    one expects the following (E)UV signatures of fundamental trapped FSMs.
Firstly,  the intensity variations consistently possess
    the same period as the FSM.
Secondly, the same is true for the variations
    in the Doppler shift, which nonetheless 
     can be seen only when the loop is sampled at non-$90^\circ$ angles with respect to its axis.
    In general, a non-$90^\circ$ phase-difference exists between
        the intensity and Doppler shift profiles, with the deviation from the $90^\circ$
        expectation from non-EI effects. 
Thirdly, in general the variations in the Doppler width possess
    two periodicities, one at the wave period ($P$) and the other at $P/2$. 
    This behavior results from 
       the competition between the broadening due to bulk flow
        (the $P/2$-periodicity) 
       and that due to temperature variations (the $P$-periodicity), 
       meaning that the relative importance of the two periodicities
       may be considerably different for different emission lines.
For non-flaring conditions, the effect due to bulk flow perturbations
    dominates over that due to fluctuating
    thermal motions, leading to a dominant $P/2$-periodicity.
The opposite happens for flaring conditions,
    resulting in a dominant $P$-periodicity.       
These signatures are in close agreement
   with the IRIS flare line Fe~\MyRoman{21}~1354~\AA\ 
   observations reported by 
   \citet{2016ApJ...823L..16T}, with some indication of
   the weaker $P/2$-periodicity
   discernible in Figure~4 therein despite the cadence issue. 
This agreement lends some concrete support to the
   interpretation of this IRIS observation
   in terms of a fundamental FSM
   (also see Table~2 in \cite{2013A&A...555A..74A} 
   and Fig.~15 in \cite{2019PASJ...71R...1H} for 
   a summarizing diagram of observables in non-flaring conditions).
       
The detailed signatures aside, the point to draw from this series of 
   forward modeling studies on fundamental FSMs
   is that the spectroscopic variations
   are in general detectable when coronal loops are not too poorly resolved
   spatially.
For fundamental FSMs, the periodicity found from the intensity variations
   faithfully reflects the wave period for trapped and leaky modes alike.
This latter point was shown by \citet{2019ApJ...883..196S}, 
   who also demonstrated that the damping time derived
   from the intensity variations can largely reflect
   the wave damping time if the loop temperature is
   not drastically different from the nominal line formation temperatures
   expected with equilibrium ionization.      
    
\subsection{Non-Thermal Emissions}
\label{sec_forward_nonthermal}    

QPPs are frequently detected
   in radio and microwave emissions of solar flares. 
However, telling their physical cause(s) 
   and hence their connections to FSMs
   is non-trivial. 
Basically, all radio emission mechanisms can be divided
   into incoherent (say, gyrosynchrotron, free-free) 
   and coherent (maser and plasma) ones, 
   and interpreting and modeling the QPPs in radio bursts produced by different mechanisms require different approaches. 
As a rule, the incoherent gyrosynchrotron mechanism is 
   responsible for the so-called microwave flaring continuum, while the coherent plasma mechanism is believed to produce the decimetric and metric radio bursts, including both the broadband continuum and various fine temporal and spectral structures in these ranges.
	
\subsubsection{Gyrosynchrotron Microwave Emissions}
In the microwave range ($\sim 3-100$ GHz),     
   the nonthermal flare emissions are produced mainly
   by the incoherent gyrosynchrotron mechanism 
   \citep{1968Ap&SS...2..171M, 1969ApJ...158..753R}.
This mechanism is highly sensitive to the parameters
   characterizing the magnetic field, energetic electrons,
   and thermal plasmas.
As such, the observed QPPs can result from
   both direct modulations of the emission source parameters by MHD waves
   and quasi-periodic injections of energetic electrons. 
Since the basic characteristics of QPPs, the periods for instance,
   can be interpreted in multiple ways, identifying their origin requires looking for more subtle features (e.g., relations between the pulsations in different spectral channels) and comparing observations with simulations. 
In particular, a typical solar microwave gyrosynchrotron spectrum peaks
   at around $\sim 10$ GHz and consists of both
   optically thick (below the peak frequency)
   and optically thin (above the peak frequency) parts, thereby
   implying different dependencies on the source parameters
   and hence, possibly, different oscillation patterns.
In a sufficiently dense plasma, the spectral shape
   can be additionally affected by the so-called Razin effect, 
   which suppresses the gyrosynchrotron emission at low frequencies.

Relatively simple models were adopted in
   the first simulations of gyrosynchrotron
   microwave emissions from oscillating flaring loops 
   \citep{2002AstL...28..783K, 2006A&A...446.1151N, 2007ARep...51..588R, 2008ApJ...684.1433F, 2012ApJ...748..140M}.
By simple, we mean that the simulations largely
   assumed a homogeneous emission source, 
   neglected the spatial structures of the oscillations,
   and retained only the temporal variations of the source parameters that are typical of the considered MHD mode. 
Despite this, when invoking standing FSMs, this approach was able to capture 
   the essential consequences of the oscillating axial magnetic field ($B_{\parallel}$),
   the defining characteristic of FSMs. 
For instance, concentrations of thermal and nonthermal electrons
   were allowed to depend on the time-varying $B_{\parallel}$.
Likewise, the energies of non-thermal particles were allowed to fluctuate
   as happens in a time-varying magnetic trap \citep{2012ApJ...748..140M}. 
As an exemplary result from the first simulations, 
   the commonly observed coherent behavior of microwave QPPs
   in a wide range of frequencies were
   interpreted as an indication of high plasma densities
   in flaring loops, 
   when the gyrosynchrotron emission is strongly affected by the Razin effect.

\begin{figure}
\centering
\includegraphics[width=0.9\columnwidth]{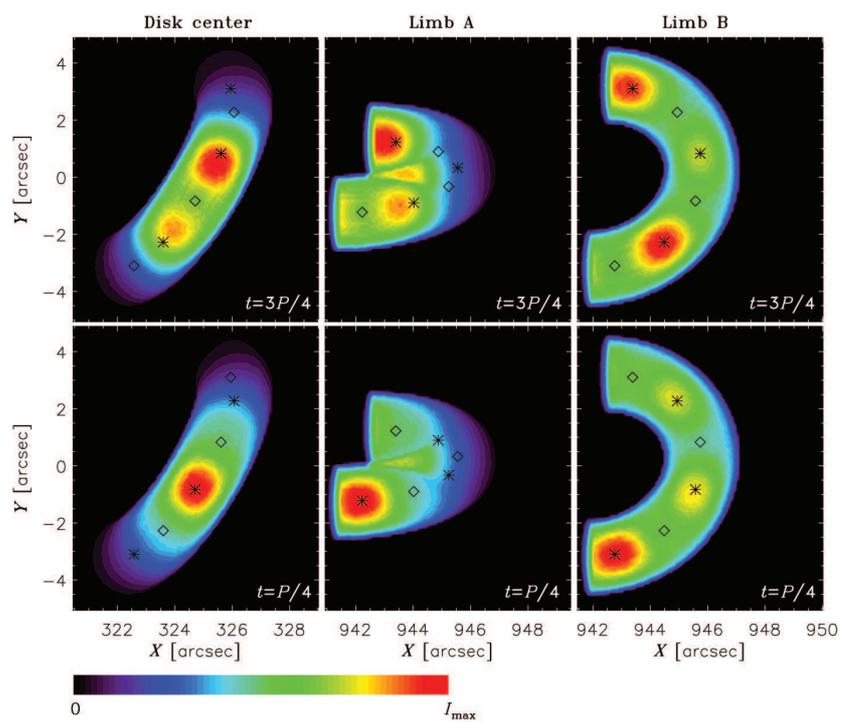}
 \caption{
Two-dimensional maps of the microwave emission
  from a curved loop containing three standing sausage waves, 
  for three different loop orientations
  and two oscillation phases. 
The asterisks ($\ast$) and diamonds ($\diamond$) 
  mark the points with the maximum / minimum magnetic field at the loop axis, respectively..
Taken from \citet{2015SoPh..290.1173K}  with permission.    
}
 \label{FigLoopSausage}
\end{figure}

Later simulations, however, have revealed that the
   spatial structures of MHD waves need to be taken into account.
\citet{2014ApJ...785...86R, 2015A&A...575A..47R}
and \citet{2015SoPh..290.1173K} 
   adopted the well-known ER83 model of MHD oscillations in an over-dense cylinder, 
   thereby accounting for the parameter variations
   both along and across the cylinder, and incorporating
   both the axial and radial components of the magnetic field.
The concentration of nonthermal electrons inside the cylinder
   was assumed to be proportional to the thermal plasma density
   \footnote{The fluctuating plasma velocity was not considered because it does not affect the gyrosynchrotron emission.}. 
With this source model, computing the gyrosynchrotron emission
   required a full 3D approach, for which purpose
   the Fast Gyrosynchrotron Codes \citep{2010ApJ...721.1127F} were used.
\citet{2014ApJ...785...86R, 2015A&A...575A..47R} considered
    a straight cylinder, which could be observed from different angles. 
\citet{2015SoPh..290.1173K} considered a curved semi-circular loop
    (see Figure \ref{FigLoopSausage}) where the viewing angle varies continuously along the loop. 
The following main results have been obtained:
\begin{itemize}
\item
If the thermal plasma density is relatively low such that the Razin effect is negligible, 
   then the high-frequency (optically thin) and low-frequency (optically thick) emissions oscillate in phase.
An exception occurs in a relatively narrow band just below the spectral peak, 
   and the oscillations therein are shifted by roughly a quarter of
   an MHD wavelength with respect to those at higher and lower frequencies.
\item
If the thermal plasma density is high, 
   then the emission spectrum can be dominated by the Razin effect. 
In this case, the emissions across almost the entire spectrum oscillate
   in phase.
An exception occurs at the lowest frequencies (much lower than the spectral peak). 
The oscillation phase therein reverses, and
   the reversal frequency depends on the viewing angle.
\item
The modulation depth of microwave oscillations is higher
   in the optically thin frequency range, reaching
   a few tens percent for reasonable parameters
   of nonthermal particles and MHD waves. 
\item
In all cases and at all frequencies, 
   the emission polarization (the absolute value of Stokes $V$) oscillates in phase with the intensity (Stokes $I$).
\item
Averaging over the visible source area
   (for both the optically thick and thin emissions)
    or along the line of sight (for the optically thin emission) 
    reduces the amplitudes of the microwave oscillations, 
    which limits our ability to detect and interpret them.
\end{itemize}

In solar flares, the observed microwave QPPs are usually broadband,
    meaning that the pulsations are synchronous at all frequencies
    including the optically thin and thick ranges \citep[e.g.,][]{2007ARep...51..588R, 2008ApJ...684.1433F, 2008A&A...487.1147I, 2012ApJ...748..140M}. 
According to the above-referenced simulations, 
    such a behavior is well consistent with the modulation of the emission by FSMs
    for a broad range of parameters (both low and high plasma densities). 
However, the same pattern is also expected for
    the quasi-periodic injection of energetic particles. 
It then follows that a reliable identification of the physical causes of QPPs
    requires using observations in other spectral ranges (X-ray and EUV, to name but two) and/or imaging spectroscopy microwave observations
    with sufficient spatial and spectral resolutions.

\subsubsection{Oscillations in Zebra Patterns}
Zebra patterns are an intriguing type of fine structures
    in the dynamic spectra of solar radio emissions observed in metric, decimetric and microwave ranges.
They appear as sets of nearly parallel bright and dark stripes
    superimposed on broadband type IV bursts. 
The most common interpretation links the formation
    of zebra patterns to the so-called double resonance effect \citep{1975SoPh...44..461Z, 2007SoPh..241..127K}. 
According to this model, individual zebra stripes are formed
    at the locations where the specific resonance condition is satisfied:
\begin{equation}\label{DPR}
f_{\mathrm{uh}}=\sqrt{f_{\mathrm{pe}}^2+f_{\mathrm{ce}}^2}\simeq sf_{\mathrm{ce}},
\end{equation}
   where $f_{\mathrm{pe}}$, $f_{\mathrm{ce}}$ and $f_{\mathrm{uh}}$ 
   are the electron plasma frequency, electron cyclotron frequency and upper-hybrid frequency, respectively.
Furthermore, $s$ is an integer cyclotron harmonic number (usually $s\gg 1$). 
The radio emission itself is produced at the upper-hybrid frequency
   or its second harmonic, due to the nonlinear transformation of plasma waves excited by nonthermal electrons with a loss-cone distribution. 
In an inhomogeneous coronal magnetic loop
   (where the ratio $f_{\mathrm{pe}}/f_{\mathrm{ce}}$ is variable), 
   condition (\ref{DPR}) is satisfied at different heights for different harmonic numbers, which results in the formation of the characteristic striped spectrum. 
Since an MHD oscillation affects the local values of the magnetic field
   and/or plasma density (and hence $f_{\mathrm{ce}}$ and/or $f_{\mathrm{pe}}$), 
   the locations of the double plasma resonance layers are changed accordingly, which should result in periodic oscillations (``wiggles'') of the zebra stripes. 
These wiggles were indeed observed
   in many events \citep[e.g.,][]{2013ApJ...777..159Y, 2018ApJ...855L..29K, 2020A&A...638A..22K}.
The observed periods of oscillations ($\sim 0.5-2$ s) are
    consistent with the periods of FSMs, 
    and the amplitude of wiggles ($\Delta f/f\sim 1-2\%$) implies a relatively low amplitude of the modulating MHD waves.

\citet{2016ApJ...826...78Y} simulated the modulation of zebra stripes
    by fast sausage oscillations. 
The emission source (in the equilibrium state) was modeled by a straight
    field-aligned overdense plasma slab \citep{1982SoPh...76..239E} with a constant magnetic field but gradually varying (along the slab) plasma density. 
After an initial perturbation, the evolution of the plasma and magnetic
    field parameters was simulated numerically using the Lare2D code \citep{2001JCoPh.171..151A}.
Both standing and propagating fast magnetoacoustic waves were considered. 
The time-dependent locations of the double plasma resonance layers
    and corresponding emission frequencies were computed 
    assuming that the emission occurs at the slab axis. 
The simulations demonstrated that the MHD waves 
   (with the fluctuation amplitudes of $\Delta B/B_0\sim 0.01-0.02$
    and $\Delta \rho/\rho_0\sim 0.02-0.04$) are able to produce wiggles similar to the observed ones.
Furthermore, standing MHD waves produce synchronous wiggles
    in adjacent zebra stripes, 
    while a propagating wave leads to
    a delay (phase shift) between the wiggles in different stripes.

\citet{2013ApJ...777..159Y} concluded that the quasi-periodic wiggles
   of the zebra pattern observed on 2006 December 13 were caused by a standing sausage oscillation. 
In contrast, \citet{2018ApJ...855L..29K} found that the quasi-periodic
   oscillations in the 2011 June 21 event can be interpreted in terms
   of fast sausage waves propagating upward with speeds of $3000-8000$ km $\textrm{s}^{-1}$. 
Using observations of wiggles in the zebra pattern on 1999 February 14, 
   \citet{2020A&A...638A..22K} estimated the levels of magnetic field 
   and plasma density turbulence as $\sim 10^{-3}-10^{-2}$, 
   and found indications that the turbulence was anisotropic.
The magnetic field and plasma density in this event oscillated in phase, 
   thus favoring fast waves.

\subsubsection{Fiber Bursts}
Fiber bursts (or intermediate drift bursts) are another type
   of fine spectral structures superimposed on the broadband type IV radio bursts \citep[e.g.,][]{1983ApJ...264..677B, 1998A&A...333.1034B}. 
They appear in dynamic radio spectra as narrowband frequency-drifting features, 
   often exhibit a weaker absorption stripe adjacent to the main emission stripe, 
   and usually occur in large groups (clusters). 
Two features are particularly noteworthy.   
Firstly, the characteristic frequency drift rates of fiber bursts
    are intermediate between those of type II and type III bursts. 
Secondly, the quasi-periodic time structure of fiber clusters 
    sometimes leads to the characteristic tadpole features in the wavelet spectra at a single frequency
   \citep[e.g.,][]{2011SoPh..273..393M, 2013A&A...550A...1K}.
These features    
    have prompted an interpretation that fiber bursts
    are produced due to the modulation of the background
    type IV emission by propagating fast magnetoacoustic wavetrains \citep{2006SoPh..237..153K, 2011SoPh..273..393M, 2013A&A...550A...1K}. 
On the other hand, the emission itself is believed to be produced by
    trapped nonthermal electrons with a loss-cone distribution, 
    via the plasma emission mechanism.

\citet{2006SoPh..237..153K} performed numerical simulations
   of the dynamic radio spectra modulated by a single propagating MHD disturbance superimposed on the large-scale (decreasing with height) 
   profiles of the plasma density and magnetic field. 
The small-scale plasma density and magnetic field perturbations
   were assumed to vary in anti-phase. 
The emission was assumed to be produced at the second harmonic
   of the local upper-hybrid frequency $f_{\mathrm{uh}}$. 
The radio spectrum was affected firstly by 
    the redistribution of the emission over frequency due to
    variations of the local upper-hybrid frequency
    caused by the MHD disturbance. 
Secondly, a local decrease of the magnetic field gradient
    in the MHD disturbance was assumed to
    increase the efficiency of the loss-cone
    instability and hence to increase the emission
    intensity in the corresponding region. 
Overall, the simulations demonstrated that MHD disturbances with propagation speeds
    of a few thousand $\kms$, spatial scales of a few hundred km, and amplitudes of $\Delta \rho/\rho_0\sim 10^{-3}-10^{-2}$ are able to produce fiber bursts with observed parameters, including the frequency drifts, spectra and time profiles.

\citet{2013A&A...550A...1K} performed similar simulations, 
    but considered fast magnetoacoustic wavetrains consisting of multiple pulses. 
In contrast to the model of \citet{2006SoPh..237..153K}, 
    the plasma density and magnetic field in the MHD waves were assumed to oscillate in phase, which is more adequate for the fast sausage waves.
On the other hand, possible modulation of the plasma emission mechanism
    by the MHD wave was neglected and only the emission redistribution due to the varying upper-hybrid frequency was considered. 
Again, the simulations (for waves with propagation speeds of $\sim 1000~\kms$, 
    wavelengths of $\sim 200$ km, 
    and amplitudes of $\Delta \rho/\rho_0=\Delta B/B_0=0.01$) 
    produced the dynamic spectra with the fiber bursts (and burst clusters) very similar to the observed ones. 
\citet{2013A&A...550A...1K} also performed numerical MHD simulations
    of a fast magnetoacoustic wavetrain propagating along a vertical current sheet, which allowed to reproduce the complicated time-frequency behavior (with variable frequency drifts) sometimes observed in the dynamic spectra of fiber bursts.

\subsubsection{Quasi-periodic Structures in Type IIIb Bursts}
Solar type III radio bursts often appear to comprise multiple
    narrowband slowly-drifting substructures (``striae'').
Such events are sometimes called type IIIb bursts. 
Type III bursts are produced by relativistic electron beams propagating
    along open magnetic field lines, 
    and the striae are believed to occur due to the modulation of
    the emission mechanism by small-scale inhomogeneities of plasma density.
In other words, the radio spectrum provides an instant snapshot
    of the structure of the inhomogeneities. 
Recently, one such burst (on 2015 April 16) was observed
    with unprecedented spectral and temporal resolution with the Low Frequency Array (LOFAR) in the 30-80 MHz frequency range, with the fine spectral structures studied in detail by \citet{2017NatCo...8.1515K, 2018ApJ...856...73C, 2018ApJ...861...33K, 2018SoPh..293..115S}. 
In particular, \citet{2018ApJ...861...33K} have detected
    quasi-periodic behavior in the frequency profile of the burst, 
    which suggests the presence of propagating MHD waves.

\citet{2018ApJ...861...33K} performed numerical simulations
   of the dynamic radio spectra, in application to the mentioned event. 
The emission was assumed to be produced at the local plasma frequency, 
   and the intensity variations were caused by the redistribution
   of the emission over frequency due to the fluctuations
   of the local plasma density. 
The source model included the large-scale background
   profile of the plasma density (described by the Newkirk formula) 
   perturbed by two propagating harmonic waves with variable wavelengths, phases, speeds and amplitudes. 
Comparison of the simulations with the observations (using the MCMC method) 
   revealed two quasi-oscillatory components: 
   the shorter one with a wavelength of $\sim 2$ Mm, speed of $\sim 657$ km $\textrm{s}^{-1}$ (which implies an oscillation period of $\sim 3$ s) 
   and a relative amplitude of $\Delta n/n_0\sim 0.35\%$, and the longer one with a wavelength of $\sim 12$ Mm and relative amplitude of $\sim 5.1\%$ (the speed was not reliably determined). 
These wave parameters are consistent with properties of fast magnetoacoustic wavetrains.

\section{Coronal Fast Sausage Modes with non-ER83-like Dispersion Features}
\label{sec_nonER83}    
This section is devoted to non-ER83-like equilbria, by which we mean that
   the FSMs hosted therein do not inherit the entire list of primary dispersion 
   features in section~\ref{sec_ER83_ER83like}.
As such, this section is ordered largely in accordance with that list,
   and effort is taken to make different subsections as mutually exclusive as possible. 
%However, we refrain from addressing the deviation from Feature 3 here, the purpose
%   being to focus on the situations where an unambiguous definition of
%   sausage modes is allowed.      

\subsection{Coronal Loops with Axial Flows}
\label{sec_nonER83_flow}    
For loops without equilibrium flow ($\myvec{v}_0$), 
    one is allowed to examine only a quadrant of
    the $\omega-k$ plane to examine the dispersive properties of coronal FSMs.
Such a symmetry is lost when $\myvec{v}_0$ does not vanish,
    thereby violating Feature 1
    (see the appendix of \citeauthor{2013ApJ...767..169L}~\citeyear{2013ApJ...767..169L}
    for an illustration, also the introduction therein
    for early studies addressing non-zero $\myvec{v}_0$). 
Focusing on coronal FSMs, the available studies seem to pertain exclusively 
    to a time-independent field-aligned flow in straight loops, 
    although the transverse profiles were allowed to be 
    either step~\citep[e.g.,][]{2016RAA....16...92Y}
    or continuous~\citep[e.g.,][]{2014A&A...568A..31L}.
Of interest were standing modes in loops of length $L$,
    for which the axial wavenumbers ($k^{+}$ and $k^{-}$)
    of the component running waves need to satisfy $k^{+}+k^{-} = 2\pi n/L$
    for the transverse displacements at the loop border 
    to be identically zero on the bounding photospheres. 
Here $k^{+}$ and $k^{-}$ are assumed to be positive, the superscripts $\pm$
    refer to the directions of axial propagation, and $n=1, 2, \cdots$ represents   
    the axial harmonic number. 
Relative to the static case,     
    a finite $\myvec{v}_0$ may considerably lower the maximum
    length-to-radius ratio 
    only below which can the loop support
    a trapped standing mode. 
This may be true even for loop flows with a magnitude
    of only a few percent of the external
    \Alf\ speed~\citep[][Figure~6]{2014A&A...568A..31L}.
Seismologically, this means that one may deduce the upper limit of the loop flow strength
    if a trapped FSM is observed in a loop~\citep[e.g.,][]{2014SoPh..289.1663C}. 
However, the above-mentioned definition
    for the standing modes may prove overly restrictive.
The reason is, the fluid motions on the bounding dense layers 
    do not need to strictly follow the required pattern, given that
    coronal loops are usually only in quasi-equilibrium
    \citep[see e.g.,][for an illustration]{2008A&A...488..757G}. 

\subsection{Coronal Loops with Axial Stratification}
\label{sec_nonER83_zInhom}    
Feature 2 is violated if the axial distribution of the equilibrium parameters
    is no longer uniform.
Despite that, in general
    the axial fundamental and its harmonics can still be unambiguously
    defined for standing modes with the number of anti-nodes along the axial direction.
Focusing on FSMs, so far the theoretical studies addressing this aspect
    exclusively adopted a straight configuration, while
    the axial inhomogeneity is allowed to be either
    in the magnetic field strength 
    or in the density. 
The former inhomogeneity results in a nonuniform axial distribution 
    of the loop cross-sectional area.
In the slab geometry mimicking flare loops, 
    \citet{2009A&A...494.1119P} showed 
    that this inhomogeneity has only a weak influence on
    the periods of trapped FSMs when the slab half-width
    expands by up to $\sim 50\%$ from footpoint to apex
    (e.g., Figures 6 and 8 therein).
The periods in expanding slabs were found to differ only slightly from the cases
    with uniform cross-sections, provided that the minimal half-width in
    the former case is adopted to evaluate the periods in the latter.
Likewise, in the cylindrical geometry, 
    \citet{2018JPhA...51b5501C} found that 
    a non-uniform density distribution in the axial direction
    has little influence on the periods and damping times
    of leaky FSMs in AR loops, even if
    the loop height exceeds the gravitational scaleheight by a factor of three
    (Figure~7 therein).
With this study {  and the pertinent studies on magnetic arcades
    \citep{2007A&A...471..999D,2007A&A...476..359D}
    },
    we note that although the axial inhomogeneity in density
    is usually attributed to gravitational stratification, 
    no eigenmode analysis is available that explicitly
    incorporates buoyancy in the restoring forces for coronal FSMs.

\subsection{Coronal Loops with Elliptic Cross-sections}
\label{sec_nonER83_ellip}    
This subsection considers an equilibrium that differs from the ER83 setup
    only in replacing the circular cross-section therein with 
    an elliptic one. 
While apparently simple, this replacement destroys the perfect azimuthal symmetry
    that enables the original classification of the collective modes
    in terms of the azimuthal wavenumber (see Equation~\ref{eq_Fourier_ansatz}).
Initiated by~\citet{2003A&A...409..287R},     
    the theoretical studies on the modes in this configuration proved much more involved,
    and were exclusively done with the elliptic coordinate system~\citep[see e.g.,][for follow-up studies]{2009A&A...494..295E,2011A&A...527A..53M}.
An elliptic loop boundary is specified by the major and 
    minor half-axes ($a$ and $b$), from which a parameter 
    is defined as $\sigma = \sqrt{a^2 - b^2}$. 
With this definition, we note that the only study that detailed sausage modes was 
    offered by \citet[][EM09 hereafter]{2009A&A...494..295E}, and 
    the dispersion relations (DRs) derived therein were complicated to such an extent
    that numerical solutions to the DRs were provided only for the situations
    where $|\mu_{\rm i, e}^2 \sigma^2/4| \ll 1$.
Note that the definitions for $\mu_{\rm i, e}^2$ were given in
    Equation~\eqref{eq_def_mu}, and EM09 examined only
    trapped modes ($\mu_{\rm e}^2 <0$).
Collective modes associated with breathing motions of the loop boundary
    were shown to exist, enabling them to be classified as sausage modes.
Qualitatively, the dispersion curves for both fast and slow sausage modes 
    are in close agreement with the ER83 results, evidenced by a comparison
    between e.g., Figures~3b and 5 in EM09 with Figure~\ref{fig_DR_tophat_vph} here.
Quantitatively, the periods of trapped FSMs differ little
    from the circular-cross-section cases even for an $a/b$ as large as
    $\sim 2.2$ (Figure~6 in EM09). 
The result is that, Feature 3 of FSMs is likely to be preserved
    rather than being violated 
    in the equilibria with circular cross-sections.
This makes the present subsection somehow different from     
    the rest,
    the purpose being to show that a loss of perfect azimuthal symmetry does not necessarily
    makes sausage modes indistinguishable. 

\subsection{Coronal Loops with Continuous Transverse Density Structuring}
\label{sec_nonER83_contTransProf}    
\begin{figure}
\centering
\includegraphics[width=0.6\columnwidth]{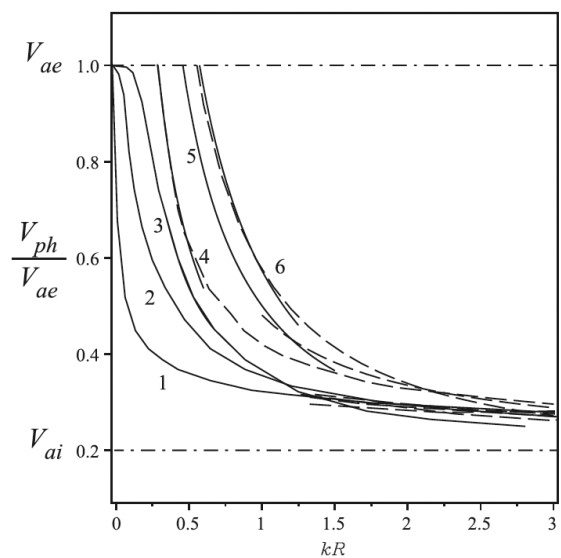}
 \caption{
 Dependence on the axial wavenumber $k$ of the axial phase speed ($\vph$)
     for fast sausage modes in a zero-$\beta$ cylindrical equilibrium.
 The equilibrium magnetic field is uniform, and a continuous density profile is
     realized with $f(r)$ as in Equation~\eqref{eq_rho_profile_general}.
 Curves 1 to 3 pertain to an $f(r)$ of $1/[1+(r/R)^\alpha]$ with 
     $\alpha = 1/2, 1, 3/2$.
 Curves 4 to 6 pertain to an $f(r)$ of $\exp[-(r/R)^\alpha]$ with 
    $\alpha =1, 2, 50$.
 The density contrast $\rhoi/\rhoe = 25$, and 
    only transverse fundamentals are considered.   
 Cutoff wavenumbers are present in curves 4 to 6 
    but absent in curves 1 to 3.
 Taken from \citet{2015ApJ...810...87L}  with permission.    
}
 \label{fig_LN15}
\end{figure}

This subsection addresses those equilibria for which either Feature 4 or Feature 5
    (or both) is absent for the hosted FSMs.
Replacing the step transverse density profile with a continuous one is one possibility. 
This has been partially addressed in Section~\ref{sec_ER83_ER83like}, 
    and the zero-$\beta$ assumption
    and the generic density profile $f(r)$
    (Equation~\ref{eq_rho_profile_general})
    remain relevant here. 
Figure~\ref{fig_LN15}, taken from \citet{2015ApJ...810...87L}, shows the dependence of the
    axial phase speed ($\vph$) on the axial wavenumber $k$ 
    of the transverse fundamental FSMs
    for two profiles,
    one being $f(r) = 1/[1+(r/R)^\alpha]$ 
    (curves 1 to 3, corresponding to $\alpha = 1/2, 1, 3/2$)
    and the other being $f(r) = \exp[-(r/R)^\alpha]$
    (curves 4 to 6, corresponding to $\alpha =1, 2, 50$).
One sees that cutoff wavenumbers are present in curves 4 to 6 
    but not in curves 1 to 3, thereby explicitly demonstrating that
    $f(r)$ needs to decrease more rapidly than $r^{-2}$ at large distances 
    for cutoff wavenumbers to exist.
For both cylindrical and slab geometries, 
    \citet{2018ApJ...855...53L} further showed that for arbitrary $f(r)$,
    the cutoff wavenumbers can always be expressed by $\kcl R = d_l/\sqrt{\rhoi/\rhoe-1}$ with $d_l$ being some constant.
When $f(r)$ decreases more (less) rapidly than $r^{-2}$ 
    at large $r$, this $d_l$ is finite and $l$-dependent (vanishes regardless of $l$).
When $f(r) \propto r^{-2}$ at large $r$, 
    this $d_l$ is finite but independent of $l$.
    
Regarding Feature 5, it was already said that 
    the monotonicity of the $\vgr-\omega$ curves of trapped FSMs 
    is largely determined by 
    the transverse density gradient at the loop axis.
Recall that in general $f(r) \approx 1-(r/R)^\nu$ close to the axis, 
    with $\nu$ some steepness parameter.
Recall further that $\vgr$ for trapped FSMs always decreases from $\vae$ 
    with increasing $k$, whether cutoff wavenumbers exist
    \citep[e.g.,][]{2014ApJ...781...92V,2019ApJ...886..112K} or not
    \citep[e.g.,][]{2017ApJ...836....1Y}.  
As shown by \citet{2018ApJ...855...53L}, 
    $\vgr$ for $kR \gg 1$ can be approximated by 
    $\vgr^2/\vai^2 \approx 1+(1-\zeta)(c_l/kR)^\zeta$, where
    $c_l$ is some constant that depends on $l$ and $\rhoi/\rhoe$. 
The point is that, $\zeta$ is given by $2\nu/(2+\nu)$.
When $\nu >2$, one finds that $\zeta >1$ and therefore the $\vgr-\omega$ curves are
    non-monotonic because $\vgr$ approaches $\vai$ from below at large $k$ or equivalently large $\omega$.
Likewise, $\vgr$ approaches $\vai$ from above at large $\omega$ when     
    $\nu < 2$, and the $\vgr-\omega$ curves
    are likely to be monotonic. 
In these two situations, one is allowed to ``predict''
    the monotonicity by using only $\nu$ out of the full specification
    of $f(r)$.
The full specification, however, is necessary for one to deduce the 
    monotonicity when $\nu=2$.

\subsection{Coronal Loops with Magnetic Twist}
\label{sec_nonER83_twist}    
\begin{figure}
\centering
\includegraphics[width=0.6\columnwidth]{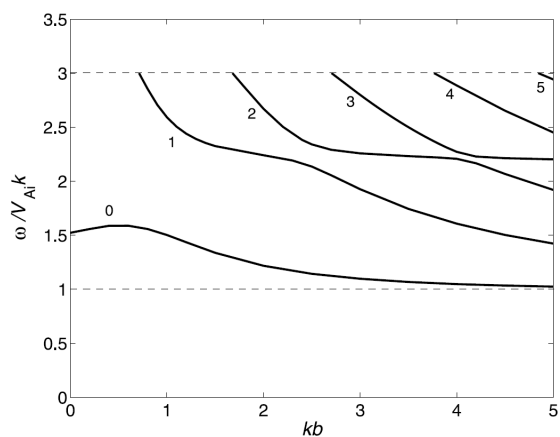}
 \caption{
 Dependence on the axial wavenumber $k$ of the axial phase speed ($\omega/k$)
     for fast sausage modes in a zero-$\beta$ cylindrical equilibrium.
 The equilibrium magnetic field $\myvec{B}_0$ is uniform and axially directed in the regions
     $r<b$ and $r>a$, whereas $\myvec{B}_0$ is purely azimuthal
     in the annulus $a < r <b$.      
 The mass density is distributed in a such a way that the \Alf\ speed determined
     by $\myvec{B}_0$ is piece-wise constant across $r=b$.       
 Note that the transverse fundamental is trapped regardless of $k$.
 Taken from \citet{2012SoPh..280..153K} with permission.    
}
 \label{fig_KMR12}
\end{figure}

This subsection addresses the equilibria supporting FSMs that do not possess
   Feature 4, and consequently may not inherit Feature 6.
Introducing magnetic twist in the cylindrical geometry
   $(r, \theta, z)$, namely allowing the equilibrium magnetic field
   $\myvec{B}_0$ to possess an azimuthal component, is one possibility.
This twist may be introduced in the interior
   \citep[e.g.,][]{2007SoPh..246..101E},    
   the exterior \citep[e.g.,][]{2018JASTP.175...49L,2019ApJ...882..134L},
   an annulus between the two \citep[e.g.,][]{2012SoPh..280..153K},
   or throughout the equilibrium~\citep[e.g.,][]{2015ApJ...810...53G,2016ApJ...823...71G}.
If the ER83 equilibrium is modified only by revising $\myvec{B}_0$ in the interior such that
   its $\theta$-component $B_{0\theta}(r)\propto r$, 
   then \citet{2007SoPh..246..101E} showed that the dispersive properties of trapped FSMs
   are hardly affected even when the maximum of $B_{0\theta}$ reaches $20\%$ of the 
   axial magnetic field strength ($B_{0z}$) at the loop axis (Figure~10 therein).
However, this is no longer true if some external twist is introduced, even if
   the mass  density remains piece-wise constant and the axial component of $\myvec{B}_0$
   is essentially piece-wise constant~\citep{2015ApJ...810...53G}.
In this case, at least the transverse fundamental becomes trapped for arbitrary 
   axial wavenumber $k$, provided that the magnetic twist is not extremely weak
   (Figure 7 therein).   
The same happens for the zero-$\beta$ equilibrium where the magnetic twist is introduced
   only in an annulus $b < r < a$, with $\myvec{B}_0$ being purely azimuthal and
   {  proportional to $1/r$ therein}
   \citep{2012SoPh..280..153K}.
For mathematical convenience, the density was specified such that the \Alf\ speed
   as determined by $B_0$ is piece-wise constant across $r=b$. 
In this case no cutoff wavenumber exists for the transverse fundamental, for which
   $\omega$ is approximately
\begin{equation}
\displaystyle 
\omega^2 \approx k^2 \vae^2 \frac{\ln(a/b)}{2+\ln(a/b)}
\label{eq_Khog_29}
\end{equation}    
   for $kb \ll 1$. 
This happens despite that cutoff wavenumbers persist for the rest of transverse harmonics, 
   as seen in Figure~\ref{fig_KMR12} which pertains to $a/b = 2$ and $\vae/\vai = 3$
   (taken from \citeauthor{2012SoPh..280..153K}~\citeyear{2012SoPh..280..153K}).

When magnetic twist is present, trapped FSMs will in general resonantly couple 
    to the $m=0$ \Alf\ waves when the equilibrium parameters are allowed to
    vary continuously~\citep[see e.g., the review by][and references therein]{2011SSRv..158..289G}. 
Indeed, the resonant absorption of trapped FSMs in the \Alf\ continuum sets in when
    \citet{2016ApJ...823...71G} replaced the step density profile in \citet{2015ApJ...810...53G} with one that varies in a transition layer (TL)
    continuously connecting the internal and external densities.
For density contrasts representative of AR loops,
    two sets of trapped FSMs are allowed at small $k$, one with
    axial phase speeds $\vph$ comparable to the external \Alf\ speed
    whereas $\vph$ for the other is close to the internal one.
The former tends to resonantly damp so rapidly
    that it may not survive a period, whereas the latter has much larger a 
    chance to survive several periods unless the magnetic twist is too strong
    (Figure~6 therein).

\section{Seismology}
\label{sec_nonER83_seis}    
In principle, any theoretical analysis of coronal FSMs
    can be of seismological use. 
However, any theoretical progress inevitably involves 
    more parameters than contained in ER83. 
With the ER83 equilibrium apparently idealized,
    these additional parameters should in principle reflect
    steps toward reality to justify their introduction.
We take the considerations of continuous transverse structuring,
    axial stratification, and loop curvature as self-evident.      
The introduction of the rest is usually justifiable as well.
Take axial flows for instance.
Subsonic flows have been shown to be 
    ubiquitous in the solar atmosphere~\citep[e.g.,][]{2004psci.book.....A}
    and have been found in oscillating structures~\citep[e.g.,][]{2008A&A...482L...9O, 2008MNRAS.388.1899S}.
%Although less frequent, flow speeds reaching the Alfv\'{e}nic range
%   have been seen associated with explosive events~\citep[e.g.,][]{2003SoPh..217..267I,2005A&A...438.1099H}.
Likewise, the ubiquity of magnetic twist in the solar atmosphere is evidenced
   by observations of, say, spicules~\citep[e.g.,][]{2012ApJ...752L..12D} and
   tornadoes~\citep[e.g.,][]{2012Natur.486..505W, 2012ApJ...752L..22L}.
In particular, magnetic twists in coronal loops may result
   from the ubiquitous rotating network magnetic fields~\citep[e.g.,][]{2011ApJ...741L...7Z},
   or ultimately from the emergence of flux tubes
   out of the convection zone~\citep[e.g.,][]{2011SoPh..270...45L}.

That said, the observational instances of 
   candidate coronal FSMs remain less extensive than kink modes.
They largely pertain to second-scale QPPs in solar flares, be them
   spatially unresolved
   (see e.g., the list compiled in Table~1 of 
   \citeauthor{2004ApJ...600..458A}~\citeyear{2004ApJ...600..458A})   
   or resolved
   (for radio observations, see e.g., 
   \citeauthor{2003A&A...412L...7N}~\citeyear{2003A&A...412L...7N};
   for UV and EUV observations, see e.g.,
   \citeauthor{2012ApJ...755..113S}~\citeyear{2012ApJ...755..113S},
   \citeauthor{2016ApJ...823L..16T}~\citeyear{2016ApJ...823L..16T};
   also the review by Zimovets et al. in this issue).
As has been discussed, one primary reason for observers to look for candidate FSMs
   by examining this particular timescale is the persistence of Feature~4 for ER83-like 
   equilibria, namely the existence of cutoff
   wavenumbers~(see Equation~\ref{eq_tophat_Ptau_k0}).
Resolving this short periodicity is readily achievable with radio instruments 
   like NoRH~\citep{1997LNP...483..183T},
   but is only marginally possible in (E)UV
       even with high-cadence instruments
       like IRIS~\citep{2016ApJ...823L..16T}.
High-cadence measurements in white light and coronal forbidden lines
   are available with ground-based instruments like the Solar Eclipse Corona Imaging System~\citep[SECIS,][]{2001MNRAS.326..428W},
   but were exclusively performed at total eclipses.    
With the cadence issue in mind, we proceed to discuss the seismological applications
    that can be enabled by recent theoretical findings. 
For simplicity, we will restrict ourselves to zero-$\beta$ studies, and 
    consider only transverse fundamentals.
While higher transverse harmonics cannot be ruled out~\citep[e.g.,][]{2005A&A...439..727M},
    to say anything definitive would require the transverse profiles
    of the pertinent eigen-functions to be spatially resolved, 
    which proves demanding even for the transverse fundamental.

\subsection{Standing Modes}
Let us start by noting that usually three steps are involved in seismological
    applications of an observed oscillatory signal.
Step 1, one experiments with different physical interpretations by contrasting
    measurables with theoretical expectations, with periods (and damping times)
    often of primary importance.
Step 2, with a chosen interpretation for a given theoretical framework, one
    inverts measurables for unknowns, with the \Alf\ time often topping the list.
Step 3, one assesses whether the derived unknowns are reasonable, 
    with either additional physical considerations or common sense.
    
Going through these steps helps appreciate the advantage of 
    the violation of Feature 4, for which purpose we consider
    spatially resolved observations such that the loop length ($L$)
    and radius ($R$) are known. 
In addition, focus for now on axial fundamentals in the trapped regime.
In the ER83 framework, the axial phase speed $\vph$ 
    in units of the internal \Alf\ speed ($\vai$)
    depends only on the density contrast $\rhoi/\rhoe$ and $L/R$.
Here $\vph = 2 L/P$ is taken as known.
With the dimensionless axial wavenumber $k R = \pi R/L$ known, 
    Equation~\eqref{eq_kcl_tophat} suggests that 
    trapped modes are allowed only when $\rhoi/\rhoe$ exceeds some critical value
    $(\rhoi/\rhoe)_{\rm crit} = [(j_{0,1}/\pi)(L/R)]^2+1 \approx 0.59(L/R)^2+1$.
%Now take the NoRH measurements reported by \citet{2003A&A...412L...7N}, 
%    where the authors practiced step~1 by associating 
%    a periodicity of $P\approx 15$~sec with an FSM hosted in a flare loop
%    with $L/R \approx 25~{\rm Mm}/3~{\rm Mm} = 8.3$.
%It then follows that $\vph = 2L/P \approx 3.3\times 10^{3}~\kms$, 
%    and $(\rhoi/\rhoe)_{\rm crit} = 0.59 (8.3)^2+1 \approx 42$.
%Evaluating the DR (Equation~\ref{eq_DR_tophat}) indicates that $\vph/\vai$ for
%    this given $kR$ depends very weakly on $\rhoi/\rhoe$ even when
%    it reaches $1000$.
%Hence $\vph/\vai = \sqrt{(\rhoi/\rhoe)_{\rm crit}}  \approx 6.5$ regardless of
%    the unknown $\rhoi/\rhoe$.
%This practice of step 2 then yields that $\vai \approx 510~\kms$. 
%Note that the external \Alf\ speed  
Now consider the NoRH measurements reported by \citet{2010SoPh..267..329K}, 
    who found a periodicity of $P\approx 30-40$~sec in connection with
    a flare loop with $L \approx 25~{\rm Mm}$ on 21 May 2004.
This event was employed by \citet[][hereafter KMR12]{2012SoPh..280..153K}
    to illustrate the limitations
    of the applicability of the ER83 equilibria, and we will largely follow their
    reasoning by further assuming that $R = 2.5$~Mm and taking $P=40$~sec. 
As step 1, we assume that this periodicity pertains to an axial fundamental FSM
    in the trapped regime.
To initiate step 2, we note that 
    $\vph = 2L/P \approx 1250~\kms$, 
        and $(\rhoi/\rhoe)_{\rm crit} \approx 0.59 (10)^2+1 =60$.
Evaluating the DR (Equation~\ref{eq_DR_tophat}) indicates that $\vph/\vai$ for
    this given $kR$ depends very weakly on $\rhoi/\rhoe$ even when
    it reaches $1000$.
Hence $\vph/\vai = \sqrt{(\rhoi/\rhoe)_{\rm crit}}  \approx 7.75$ regardless of
    the unknown $\rhoi/\rhoe$, yielding $\vai \approx 161~\kms$.
Practicing step 3, one would regard this $\vai$ to be too small.
Evidently this issue arises because $R/P$ is small, and therefore is likely to take place
    in general when one employs FSMs to interpret signals with long periods
    \footnote{For trapped FSMs, Figure~\ref{fig_DR_tophat_vph} indicates that when $\rhoi/\rhoe$ is given, $\vph/\vai$ decreases with increasing $kR$. This is not to be confused with that $\vph/\vai$ for a relatively small $kR$
    shows little dependence on $\rhoi/\rhoe$.
    When evaluating $\vph = 2L/P$, KMR12 somehow adopted twice the measured period. 
    KMR12 further supposed that $\vph$ is close to
         the external \Alf\ speed $\vae$,
         and employed the smallness of $\vae$ to 
         argue that it is difficult to reconcile the measured period
         with an FSM in an ER83 equilibrium. 
    We note that, the insensitivity of $\vph/\vai$ to $\rhoi/\rhoe$
         for a relatively small $kR$ means that $\vph/\vae$ is approximately
         $\propto \sqrt{\rhoe/\rhoi}$, and hence $\vae$ can be much larger than the measured $\vph$
         for large $\rhoi/\rhoe$.     
    Despite that, their argument remains valid.     
    }.  

KMR12 offered a possible way out, if FSMs are still adopted in step~1.
We generalize the discussions therein, and associate the inner boundary
    of the annulus ($b$) with the loop radius.     
Figure~\ref{fig_KMR12} indicates that $\vph/\vai$ depends rather weakly on $kb$ 
    when $kb \lesssim 1$ for a combination of $[a/b, \vae/\vai] = [2, 3]$.
This enables one to relate the measured $\vph$ to $\vae$ with
    Equation~\eqref{eq_Khog_29}, yielding $\vae \approx 2460~\kms$ 
    and consequently $\vai \approx 820~\kms$.
Now this value of $\vai$ becomes reasonable, and so is the density contrast
    $\rhoi/\rhoe = (\vae/\vai)^2 (a/b)^2 = 36$
    (see KMR12 for details).     
We note that one may start with interpreting
    the measured periodicity with a fast kink mode in an ER83 equilibrium,
    and still derive reasonable values for, say, $\vai$.
While a detailed forward modeling study is necessary to resolve this ambiguity,
    the point here is that the violation of Feature~4 of FSMs
    in the canonical ER83 equilibria offers more seismological possibilities. 
Nonetheless, these possibilities come at a price, which we illustrate by formally
    expressing the eigenfrequency for the transverse fundamental in the KMR12 equilibrium 
\begin{equation}
\displaystyle 
\omega  = \frac{\vai}{b}  {\cal H}\left(
    \frac{\vae}{\vai}, \frac{a}{b}, kb
    \right)~,
\label{eq_dimen_KMR12}    
\end{equation}         
    where ${\cal H}$ is some function dictated
    by the dispersion relation.
%    \footnote{Note that the equilibrium in KMR12 is fully specified
%     by five dimensional parameters $[\vai, \vae, \rhoi, a, b]$.
%     Note further that Equation~\eqref{eq_dimen_KMR12} is among
%        the many possible ways to relate the dimensionless parameters
%        constructed from the dimensional ones.
%     }. 
With $kb = \pi b/L$ and $\omega = 2\pi/P$ measurable, 
     Equation~\eqref{eq_dimen_KMR12} actually involves three unknowns
     $[\vae/\vai, a/b, \vai]$.
Solving Equation~\eqref{eq_dimen_KMR12} therefore yields
     a surface in the three-dimensional (3D) parameter space formed by the three unknowns,
     and any point on this surface is a solution to the inversion problem.
The derived $\vai$ in connection to the particular $[\vae/\vai, a/b]$
     is one example.       

That the inversion problem is usually under-determined is not specific to 
     coronal FSMs~\citep[see][for the case of kink modes]{2019A&A...622A..44A}
\footnote{
Practical seismological diagnostics using kink oscillations
  therefore includes using additional observables such as
  the shape of the damping profile
  \citep{2013A&A...551A..39H,2013A&A...551A..40P,2016A&A...589A.136P,2019FrASS...6...22P}, independent sources of information such as the EUV intensity
  \citep{2017A&A...600L...7P,2018ApJ...860...31P}, 
  and Bayesian analysis \citep{2013ApJ...769L..34A,2017A&A...600A..78P,2017A&A...607A...8P} 
  including MCMC sampling \citep{2020arXiv200505365A} which permits constraints to be 
  calculated without the need for a unique solution.}.
More importantly, this is not to say that the unknowns cannot be constrained. 
Take \citet{2015ApJ...812...22C} for instance, where the equilibria differs from ER83
     by placing a continuous transition layer (TL) between 
     a uniform cord and a uniform exterior.
In this case Feature 4 persists, and the periods and damping times for leaky modes
     can be formally expressed as
\begin{subequations}
\begin{equation}
\displaystyle 
P    = \frac{R}{\vai}  {\cal F}_P\left(
    \frac{\rhoi}{\rhoe}, \frac{l}{R}, kR
    \right)~, \label{eq_dim_Chen15_P} \\
\end{equation}
\begin{equation}
\displaystyle 
\tau = \frac{R}{\vai}  {\cal F}_\tau\left(
    \frac{\rhoi}{\rhoe}, \frac{l}{R}, kR
    \right)~, \label{eq_dim_Chen15_tau} \\
\end{equation}
\end{subequations}         
    where the functions ${\cal F}$ can be established with the dispersion relation,
    $l$ is the TL width, and $R$ the mean loop radius. 
Taking $P$, $\tau$, and $kR = \pi R/L$ as known, 
    with Equations~\eqref{eq_dim_Chen15_P} and \eqref{eq_dim_Chen15_tau} 
    one derives an inversion curve in the 3D parameter space formed by the unknowns
    $[\rhoi/\rhoe, l/R, \vai]$.
Applying this procedure to the \citet{1973SoPh...32..485M} measurements, 
    \citet{2015ApJ...812...22C} showed that the transverse \Alf\ time $R/\vai$
    is constrained to the range $[1.18, 2.13]$~sec,
    despite the diversity of the profile specifications in the TL
    \footnote{Without imaging information, this study
     assumed that $L/R \gg 1$, and drived $R/\vai$ rather than $\vai$.}.
     
The {  under-determined situation improves when, say,}
    the measured oscillatory signal involves more than one periodicity. 
These periodicities may be comparable, thereby favoring the seismological procedure
    that starts with their interpretation as involving more than one axial harmonic.
{  Much has been done to exploit coronal kink overtones, 
    with the period ratios of the axial fundamental to its first  
    harmonic ($P_1/P_2$) being particularly useful 
    \citep[see e.g., the review by][]{2009SSRv..149....3A}.
The key is, coronal kink modes tend to be observed in thin loops, 
    meaning that $P_1/2P_2$ should be close to unity if the loops are axially uniform
    given the weak dispersion. 
It then follows that any significant deviation of $P_1/2P_2$ from unity
    derives from the nonuniform distribution along the axial direction
    of the equilibrium quantities, which may be mass density \citep[e.g.,][]{2005ApJ...624L..57A} or the magnetic field strength
    \citep[e.g.,][]{2008A&A...486.1015V}, along with other possibilities.
Conversely, the measurement of $P_1/2P_2-1$ can be employed to infer
    the axial inhomogeneity lengthscale of the involved equilibrium quantity
    \citep[see e.g.,][and references therein]{2013ApJ...765L..23A}.
In the case of coronal FSMs, a significant axial inhomogeneity
    is not required for $P_1/2P_2$ to deviate significantly from unity in view of 
    the strong dispersion pertinent to both step (see Figure~1)
    and continuous transverse profiles (see Figure~6).
For simplicity, let us consider only the case where both the axial fundamental 
    and its first harmonic are trapped, and both are invoked
    in spatially resolved measurements.  
Now that the axial fundamental (its first harmonic) 
    corresponds to $kR = \pi R/L$ ($kR = 2\pi R/L$), 
    the formal expression in Equation~\eqref{eq_dim_Chen15_P}
    yields that $P_1/2P_2$ depends only on the unknown density contrast
    $\rhoi/\rhoe$ and the dimensionless transverse inhomogeneity lengthscale $l/R$.
At this point, the measured $P_1/2P_2$ is already useful to constrain the two unknowns. 
If the second harmonic (to be denoted with a subscript $3$) is additionally measured,
   then $P_1/3P_3$ can be invoked to fully constrain $[\rhoi/\rhoe, l/R]$
   for a given profile specification. 
We note that the simultaneous presence of an axial fundamental FSM together with
   its first two harmonics is not unrealistic but has been 
   reported in NoRH measurements \citep{2013SoPh..284..559K}.
In addition, the measured period ratios have been seismologically exploited
   by \citet{2015ApJ...810...87L} to yield $\rhoi/\rhoe$ and the transverse
   steepness parameter.
In this latter study, a continuous transverse distribution
   was shown to be necessary
   to account for the measurements in the first place.

The periodicities in the observed multi-periodic signals} may be disparate, 
    with their interpretation more in line 
    with modes of different physical nature. 
For example, \citet{2011ApJ...740...90V} invoked standing
    slow and fast sausage modes to interpret the multi-periodic
    signals in a flare on 8 Feb 2010 measured by PROBA2/LYRA,
    thereby deducing the plasma $\beta$ in the flare loop. 
We refrain from discussing this application here because of the assumed
    zero-$\beta$ convention.
In this context we take the NoRH measurements of a limb flare on 14 May 2013, for which
    the oscillatory signal was shown by \citet{2015A&A...574A..53K} to possess a number
    of well-defined periodicities.
Two of these are of particular interest, one with a period $P_{\rm short} \approx 15$~sec
    and a damping time $\tau_{\rm short} \approx 90$~sec, while the other corresponding to
    $P_{\rm long} \approx 100$~sec and $\tau_{\rm long} \approx 250$~sec.
These two signals were associated by the authors with a standing FSM and a standing fast 
    kink mode, respectively.
Adopting this interpretation and with the measured geometrical parameters
    ($L/R \approx 40~\megam/4~\megam$), \citet{2015ApJ...812...22C} 
    and \citet{2016SoPh..291..877G} 
    performed
    a seismology by assuming that the damping of the FSM was due to lateral leakage
    while the kink mode was resonantly damped at the \Alf\ continuum. 
A pair of expressions for $P$ and $\tau$ identical in form
    to Equations~\eqref{eq_dim_Chen15_P} and \eqref{eq_dim_Chen15_tau}
    can therefore be established for the kink mode. 
The point is that, now there are four knowns but only three unknowns for a given profile
    specification. 
In practice, \citet{2015ApJ...812...22C} neglected $P_{\rm long}$, thereby finding a unique
    inversion solution for each specification.
The theoretically expected kink period is then computed, 
    its deviation from $P_{\rm long}$ being a safety check.
The internal \Alf\ speed $\vai$ was shown to be constrained to
    a rather narrow range of $[594, 658]~\kms$.
It is noteworthy that, rather than leaving out some measurable and then performing
    a safety check a posteriori, one may handle this over-determined system 
    directly with, say, least-square minimization.
However, this intuitively appealing approach has yet to be explored
    to our knowledge.    
    
Technical details aside, underlying the seismological practice
    is that the flare loops (or coronal loops in general)
    that host FSMs possess fixed geometries and given thermodynamic conditions.
This is admittedly at odds with the dynamic nature of the solar atmosphere, and
    is particularly limiting when one considers magnetic reconnections in flares.
Indeed, the standard flare model leads to rapid apparent expansion of
    coronal loop arcades, with newly formed 
    loops constantly appearing at the top. 
In this scenario it is not unreasonable to think of
    FSMs being generated continuously in structures
    that are constantly evolving, with both continuously
    varying geometries (in particular a length increase) 
    and thermodynamic conditions. 
Assuming that wave damping occurs on a faster timescale than
    the expansion, one would then expect a continuous period drift
    in time in the detection of an apparently single sausage mode. 
Such a scenario was proposed as an explanation for the period increase
    in the high-resolution IRIS observations of a fundamental FSM
    \citep{2016ApJ...823L..16T}.            

\subsection{Impulsively Generated Sausage Wavetrains}
\label{sec_seis_impls}
As proposed by \citet{1984ApJ...279..857R}, there is rich seismic information 
    in the distinctive temporal signatures associated
    with impulsively generated sausage wavetrains (WTs hereafter).
Take the SECIS measurements acquired during the 1999 total eclipse,
    which were analyzed
    in conjunction with the simultaneous SOHO/EIT data  
    \citep{2002MNRAS.336..747W, 2003A&A...406..709K}.
A wavefront, defined by the intensity enhancement
    in the Fe \MyRoman{14} channel of SECIS,
    was found to propagate with a speed of $v_{\rm front} \sim 2100~\kms$ along
    an AR loop with a radius of $R \sim 5~\megam$~ and 
    a density contrast of $\rhoi/\rhoe \sim 2.5$. 
A periodicity of $P_{\rm front} \sim 6$~sec was detected during the passage
    of the intensity enhancement. 
These values were put to seismological practice by
    \citet[][hereafter R08]{2008IAUS..247....3R} in the framework of ER83.
R08 started by associating $P_{\rm front}$ with $P^{\rm min}$ and 
    $v_{\rm front}$ with $\vgmin$, where
    $\vgmin$ is the minimal axial group speed and 
    $P^{\rm min}$ the characteristic periodicity where 
    $\vgmin$ is attained (see Figure~\ref{fig_DR_tophat_vgr}).
This association is in line with Equation~\eqref{eq_impl_ER86}, namely
    the strongest signal tends to appear when wave packages propagating with $\vgmin$ arrive.     
R08 further assumed that $\vgmin \approx \vai$,
    and $P^{\rm min} \approx P^{\rm cutoff}$, with 
    $P^{\rm cutoff} = (2\pi/j_{0,1}) (R/\vai) \sqrt{1-\rhoe/\rhoi}$
    being the period that an FSM acquires at the cutoff. 
This $P^{\rm cutoff}$ evaluates to $\approx 2R/\vai$
   for the measured $\rhoi/\rhoe$.          
Seeing $R$ as unknown, R08 then derived with $P^{\rm cutoff}$
    an $R_{\rm seis} \approx 6.3~\megam$.
With $R$ actually measured, the R08 seismology 
    is more like a check on how safely one may adopt the physical interpretation.
We take this slightly further by
    noting that the inversion problem is over-determined.
In an ER83 equilibrium, $\vgmin/\vai$ 
    and $P^{\rm min} \vai/R$ are fully determined by $\rhoi/\rhoe$.
Taking $\rhoi/\rhoe$ as given, we therefore mean by over-determined 
   that there are two knowns
   ($\vgmin$ and $P^{\rm min}$) but only one unknown ($\vai$).
We proceed by solving the DR with the measured $\rhoi/\rhoe$, finding that
   $\vgmin = 0.887~\vai $ and $P^{\rm min} = 1.43~R/\vai$. 
With the measured $v_{\rm front}$ one then finds that $\vai \approx 2370~\kms$, 
   and derive an $R_{\rm seis} \approx 9.9~\megam$. 
That this $R_{\rm seis}$ deviates more strongly from the measured value than 
   the one derived by R08 is not surprising, because 
   $\vgmin < \vai$ and $P^{\rm min} > P^{\rm cutoff}$.
More importantly, this does not invalidate the interpretation of the measurements, 
   even if one sticks to the measured values.
%To proceed, we note that the deviation between $R_{\rm seis}$
%   and the measured value come primarily from the small values of $P^{\rm min} \vai/R$.  

Implied by R08 is that the temporal signatures of
   WTs are determined essentially by the dispersive properties
   of FSMs, which in turn depend only on the equilibrium.
However, time-dependent simulations indicated that these signatures 
   depend on the details of the initial perturbations as well,
   the temporal~\citep[][]{2019A&A...624L...4G} 
   and spatial extent~\citep[e.g.,][]{
       2015ApJ...814..135S} in particular.
This latter point was evident already in Equation~\eqref{eq_impl_Oliver25}.
Let $\sigma_z$ ($\sigma_r$) denote the extent in the $z$- ($r$)-direction. 
Likewise, let $\sigma_t$ denote the temporal extent.     
Within the ER83 framework,
   $v_{\rm front}/\vai$
   and $P_{\rm front} \vai/R$ characterizing the strongest signal in the WTs
   therefore depend on $[\rhoi/\rhoe, \sigma_t \vai/R, \sigma_z/R, \sigma_r/R]$.
Consider only $\sigma_z$ for now. 
For a given $\rhoi/\rhoe$, 
   the WTs will involve increasingly small $k$ and consequently small $\omega$
   when $\sigma_z$ increases, bringing $P_{\rm front}$ increasingly close to 
   $P^{\rm cutoff}$.
The result is that, there is no difficulty to reconcile the seismologically derived
   values with the measurements, 
   the key evidently being the flexibility associated with
   the introduction of $\sigma_z$.

The problem is, the flexibility may be overwhelming, making
   seismic information difficult to glean from observations.
The discussion so far is equivalent to seismologically deriving $\vai$ and 
   $\sigma_z$ with the measured $P_{\rm front}$ and $v_{\rm front}$.
However, the signatures of WTs depend on more parameters
   than just $\sigma_z$.      
In addition to $\sigma_t$ and $\sigma_r$, let us name but two. 
One, the distance ($h$) between the sampling point and the exciter.
While having no influence on $v_{\rm front}$ or $P_{\rm front}$,
   the signatures of WTs in general depend on $h$ as well
   (see Equation~\ref{eq_impl_Oliver25}).
Two, a dimensionless parameter ($\alpha$) characterizing the profile steepness,
   which is inevitable if one replaces the step profile
   in ER83 with a continuous one. 
Varying $\alpha$ can introduce qualitative difference to the dispersive
   properties of FSMs, and this difference can indeed be reflected in
   the temporal evolution of WTs~\citep[e.g.,][]{2017ApJ...836....1Y}.
Given the difficulty to measure $[h, \sigma_t, \sigma_r, \sigma_z]$,
   it seems rather difficult to seismologically infer the equilibrium parameters. 
Here we offer one possible way out, assuming that the timing of the exciter
   (say, a flare) is known.
The idea is that while the details of the temporal evolution of WTs and
   the corresponding Morlet spectra depend on the details of the initial perturbation,
   the spectra are in general always threaded by the $\omega - h/\vgr(\omega)$ curves.
This $\omega - h/\vgr$ profile is fully
   determined by $[R/\vai, \rhoi/\rhoe, h/R, \alpha]$.
Conversely, this set of parameters can be deduced
   if $\omega - h/\vgr$ can be observationally constructed.
We note that the inversion problem can be substantially simplified if $\rhoi/\rhoe$
   is known, which was possible at least for the above-discussed
   SECIS measurements.    
We note further that the $\omega - h/\vgr$ curve can be extracted by enhancing 
   the frequency resolution in the Morlet spectra with, say, 
   the synchrosqueezing transform~\citep[e.g.,][Figure~1c]{2018ApJ...856L..16W}.

{  Before proceeding, we note that impulsive sausage wavetrains also have 
   some bearing on the quasi-periodic fast propagating wavetrains (QFPs)
   customarily imaged with SDO/AIA.
Discovered by \citet{2011ApJ...736L..13L},
   these QFPs have been subject to extensive observational and numerical studies
   (\citeauthor{2011ApJ...740L..33O}~\citeyear{2011ApJ...740L..33O},
    \citeauthor{2012ApJ...753...53S}~\citeyear{2012ApJ...753...53S},
    \citeauthor{2016A&A...594A..96G}~\citeyear{2016A&A...594A..96G},
    \citeauthor{2017ApJ...844..149K}~\citeyear{2017ApJ...844..149K},    
    \citeauthor{2018ApJ...860...54O}~\citeyear{2018ApJ...860...54O},
    \citeauthor{2019ApJ...873...22S}~\citeyear{2019ApJ...873...22S};
    see also the review by
    \citeauthor{2014SoPh..289.3233L}~\citeyear{2014SoPh..289.3233L}).
Relevant here is what leads to the observed quasi-periodicities, 
    regarding which two possibilities arise.
One is that these quasi-periodicities derive from the source of excitation, 
    as corroborated by the coincidence of some prominent periodicities in the QFPs
    with those in the wave-generating flares
    \citep[e.g.,][]{2011ApJ...736L..13L, 2013SoPh..288..585S}.
The other is that these quasi-periodicities derive from the dispersion introduced
    by the nonuniformity of the wave-guiding funnel-like features.
First suggested in the observational study by \citet{2013A&A...554A.144Y},
    this second possibility 
    is strongly reminiscent of the evolution of impulsively
    generated sausage wavetrains 
    \citep[see e.g.,][for numerical demonstrations]{2013A&A...560A..97P,2017ApJ...847L..21P}.
In reality, the quasi-periodicities in QFPs are likely to result from
    some combination of the two possibilities.
}

\section{Summary and Prospects}
\label{sec_conc}   
Heavily involved in the establishment of coronal seismology
    \citep{1984ApJ...279..857R},
    coronal fast sausage modes (FSMs) have proven intriguing ever since
    given their dispersive properties. 
This review is therefore centered around the dispersion features of coronal FSMs 
    and the seismological applications they enable.
Overall, we conclude that      
    the canonical dispersion features found in the equilibrium examined in 
    \citet[][ER83]{1983SoPh...88..179E} 
    have been better understood physically,
    and further exploited seismologically. 
Furthermore, departures from the ER83 equilibrium may lead to qualitatively different
    dispersion features, as exemplified by the disappearance of cutoff wavenumbers
    and the resonant absorption of FSMs in the \Alf\ continuum.
Seismologically, while the interest remains largely connected to 
    quasi-periodic pulsations (QPPs) in solar flares,
    these new features broadened the range of periodicities that FSMs can be invoked to 
    account for, and consequently offered more seismological possibilities. 
That said, we list a limited number of aspects that remain to be better addressed.

\begin{enumerate}
\item The physical connection of FSMs to flare QPPs needs to be more firmly established. Instead of theoretically examining FSMs in flare loops on an individual basis,
    it will be ideal to establish some self-consistent physical or numerical model where 
    FSMs are either an outcome of or inherent ingredient in the flaring processes.
If this proves too challenging, then FSMs need to be examined in model flare loops that
    incorporate such realistic features as curvature,  
    multi-strandedness \citep{2013AstL...39..267Z},   
    significant temporal variability, 
    and strong flows \citep[e.g.][]{2011SSRv..159...19F}. 
    
\item The generation of FSMs in AR loops needs to be better understood. 
While the time signatures of rapidly running wavetrains as measured by say, SECIS,
    agree remarkably well with impulsively generated wavetrains, the exciter has yet
    to be pinpointed.  
Flow collisions could play an important role in
    thermally unstable AR loops \citep{2018ApJ...861L..15A}.    
If the generation is connected to footpoint motions in the lower solar atmosphere,
    then the coupling between the lower layers and the corona needs to be addressed.
As envisaged by \citet{1996ApJ...472..398B}, the sausage waves thus excited may
    spend a considerable amount of time in the lower layers given the low \Alf\
    speeds therein, thereby substantially increasing the periodicities.
Furthermore, while sausage modes have been observed in the
    lower layers
    \citep[e.g.,][]{2012NatCo...3.1315M,2015ApJ...806..132G,2020arXiv200711594G},
    it remains to examine whether or how they are connected to coronal FSMs.    

\item There is need for an extensive search 
    for candidate FSMs in observations of both flare QPPs
    and AR loops in view of the recent theoretical advances, 
    as was called for by \citet{2020ARAA..Nakariakov}.
    What we mean here is that attention needs not to be narrowed down to
    short periodicities inherent to the ER83-like equilibria.
    Rather, periodicities comparable to the longitudinal \Alf\ time may also
    pertain to FSMs.
    To proceed with this, however, the observational signatures of FSMs with non-ER83-like features need to be established by proper forward modeling. 
     
\item The available candidate FSMs need to be better seismologically exploited.
    The seismic information imparted by the strong dispersion of FSMs,
        often found in theoretical studies, may be too rich for seismology to be practical. 
    The high dimensionality of the parameter space
        formed by the unknowns is one primary obstacle.     
    From this perspective, probabilistic seismology within say, the Bayesian framework
        should be of considerable help~\citep{2018AdSpR..61..655A}.      
{\tiny }\end{enumerate}

\begin{acknowledgements}
{  We thank the reviewers for their constructive comments and suggestions,
    which helped improve this manuscript substantially.}	
We gratefully acknowledge ISSI-BJ for supporting the workshop on
   ``Oscillatory Processes in Solar and Stellar Coronae'', during which 
   this review was initiated. 
BL was supported by the 
    National Natural Science Foundation of China (41674172, 11761141002, 41974200).
PA acknowledges funding from his
    STFC Ernest Rutherford Fellowship (No. ST/R004285/2).    
AAK was supported by budgetary funding of Basic Research program II.16.
DJP and TVD were supported by the European Research Council (ERC) under the European Union's Horizon
    2020 research and innovation programme (grant agreement No 724326)
    and the C1 grant TRACEspace of Internal Funds KU Leuven. 
\end{acknowledgements}

\section*{Conflict of interest}
The authors declare that they have no conflict of interest.

\bibliographystyle{spbasic}      % basic style, author-year citations

\bibliography{seis_generic}

\clearpage

\clearpage

\clearpage

\clearpage

\clearpage

\clearpage

\clearpage

\end{document}